\documentclass[showpacs,pre,superscriptaddress,twocolumn]{revtex4-1}
\usepackage{amsmath}
\newcommand{\bra}[1]{\left\langle#1\right|}
\newcommand{\ket}[1]{\left|#1\right\rangle}

\newcommand{\mean}[1]{\left\langle #1\right\rangle}
\usepackage{color}

\usepackage{amsmath,amsthm, amssymb}
\usepackage{amsfonts}
\usepackage{pgf}
\usepackage{slashed}
\usepackage{graphicx}
\usepackage{dcolumn}
\usepackage{bm}
\newcommand{\overbar}[1]{\mkern 1.5mu\overline{\mkern-1.5mu#1\mkern-1.5mu}\mkern 1.5mu}
\newcommand{\kbar}{\mathchar'26\mkern-9mu k}
\newcommand{\nep}{\textrm{e}}
\newcommand{\ud}{\mathrm{d}}
\begin{document}


\title{From localization to anomalous diffusion\\in the dynamics of coupled kicked rotors}

\author{Simone Notarnicola }
\email{\texttt{email} simone.notarnicola@gmail.com}
\affiliation{%
 SISSA, Via Bonomea 265, I-34136 Trieste, Italy
}%

\author{Fernando Iemini }
\affiliation{
 Abdus Salam ICTP, Strada Costiera 11, I-34151 Trieste, Italy 
}%

\author{Davide Rossini}
\affiliation{%
Dipartimento di Fisica, Universit\`a di Pisa and INFN, Largo Pontecorvo 3, I-56127 Pisa, Italy
}%

\author{Rosario Fazio }
\affiliation{
 Abdus Salam ICTP, Strada Costiera 11, I-34151 Trieste, Italy 
}%
\affiliation{%
 NEST, Scuola Normale Superiore \& Istituto Nanoscienze-CNR, I-56126 Pisa, Italy
}

\author{Alessandro Silva}
\affiliation{%
 SISSA, Via Bonomea 265, I-34136 Trieste, Italy
}%

\author{Angelo Russomanno }%
\affiliation{%
 NEST, Scuola Normale Superiore \& Istituto Nanoscienze-CNR, I-56126 Pisa, Italy 
}%
\affiliation{
 Abdus Salam ICTP, Strada Costiera 11, I-34151 Trieste, Italy 
}%


%


\date{\today}

\begin{abstract}
We study the effect of many-body quantum interference on the dynamics of coupled periodically-kicked systems whose classical dynamics is chaotic and shows an unbounded energy increase.
We specifically focus on a $N$ coupled kicked rotors model: 
we find that the interplay of quantumness and interactions dramatically modifies the system dynamics inducing a transition between energy saturation and unbounded energy increase. We discuss this phenomenon both numerically and analytically, through a mapping onto a $N$-dimensional Anderson model. The thermodynamic limit $N\to\infty$, in particular, always shows unbounded energy growth. 
This dynamical delocalization is genuinely quantum and very different from the classical one:
using a mean field approximation we see that the system self-organizes so that the energy per site increases in time as a power law with exponent smaller than one. This wealth of phenomena is a genuine effect of quantum interference: the classical system for $N\geq 2$ always behaves ergodically with an energy per site linearly increasing in time. Our results show that quantum mechanics can deeply alter the regularity/ergodicity properties of a many body driven system.

\end{abstract}

\pacs{Valid PACS appear here}
\maketitle


\section{Introduction}
Deterministic chaos is a powerful scientific paradigm to understand the natural world~\cite{Gleick:book, Ruelle:book}. Since the first works by Lorenz~\cite{Lorenz} and May~\cite{May}, it has become suddenly clear that non-linearities in very simple maps or systems of differential equations could give rise to a complex aperiodic behaviour, strongly dependent on initial conditions. The works by Feigenbaum~\cite{Feigenbaum,Feigenbaum1} and Ruelle-Takens~\cite{Ruelle} showed that there is a universal way in which non-linear systems undergo the transition to a chaotic regime; those theories have found spectacular experimental demonstrations in the context of turbulence~\cite{Gollub,Libchaber}. The dynamics of chaotic dissipative systems in phase space converges towards sets called ``strange attractors''~\cite{eckmann,Turbo:book} whose fractal structure~\cite{Mandelbrot:book} challenges traditional geometric descriptions. Chaos is extremely pervasive and applies to fields like meteorology~\cite{Chaos_n:book}, chemistry~\cite{Turbo:book,chem:b}, economics~\cite{Kemp} and medicine~\cite{Chaos_n:book,Int,science, leon,psyco,Zapperi:book,cancer}, to the extent that even life could be thought as a chemical system self-organizing at the border between order and chaos~\cite{Kauffman:book}.

\begin{figure}
    \resizebox{\columnwidth}{!}{\includegraphics{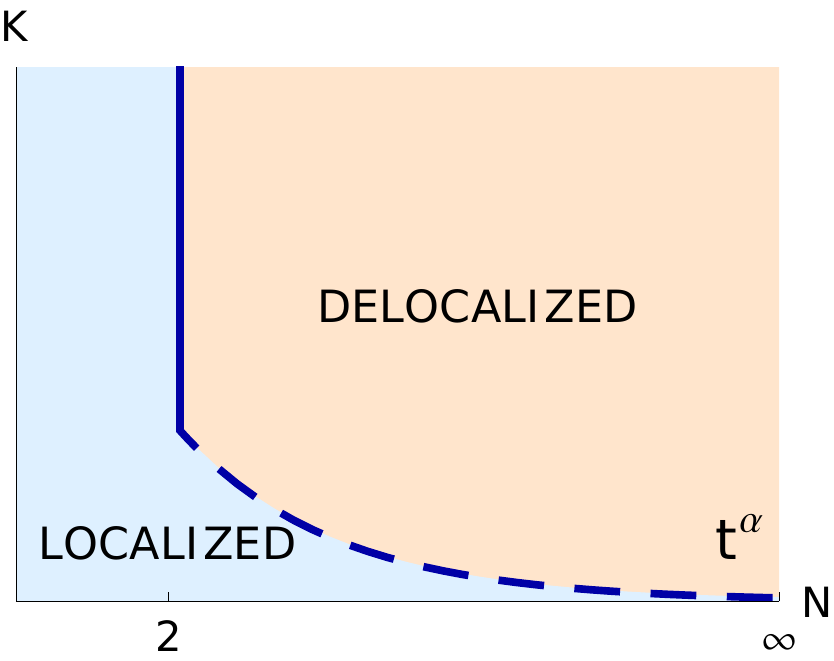}}
  \caption{\label{fig:intro} The behavior of our kicked rotors system is here sketched. For $N=1,2$ rotors the system is localized for all the values of the kick strength $K$. For $N>2$ but finite a transition occurs from localized to delocalized. For $N\rightarrow\infty$ the system is always delocalized and the mean field approach gives that the kinetic energy of the system grows subdiffusively in time ($E_t\sim t^\alpha$, with $\alpha\leq 1$).}
  \end{figure}

The focus of this work 
is the relation between chaotic dynamics from one side and ergodicity and thermalization from the other in 
quantum many-body Hamiltonian systems. This is a rather well studied topic in classical physics (see Refs.~\cite{Berry_regirr78:proceeding,Tabor:book} for a review): 
after the first studies by Poincar\'e~\cite{Berry_regirr78:proceeding}, interest in these topics was renewed by Fermi, Pasta and Ulam~\cite{Fermi_tutti} who numerically simulated a chain of non-linear oscillators, finding a complex non-thermalizing behaviour. The theoretical explanation of this fact came from Kolmogorov, Arnold and Moser~\cite{Arnold_Avez:book,Moser} who demonstrated that, for moderate non-integrable perturbations, the phase space is partly regular and partly chaotic (KAM theorem). In this intermediate situation, when there are many degrees of freedom, slow diffusion is possible in the connected chaotic cluster giving rise exponentially slow deviations from 
nearly-integrable behaviour encoded in Nekhoroshev theory~\cite{Nekhoroshev1971}. These studies are of huge theoretical importance, because they put on solid mathematical foundations the concept of ergodicity necessary for thermalization of isolated classical systems~\cite{Castiglione:book,Pettini}. If a many-body system is ergodic, all the phase space is chaotic and there is a strong dependence on initial conditions, nearby trajectories deviating from each other exponentially fast in time~\cite{Ott:book,Tabor:book}. Chaotic trajectories are extremely complex fractal objects; in the ergodic case they  uniformly fill all the available phase space~\cite{Berry_regirr78:proceeding} and time-averages over them equal the microcanonical ones: in this case thermalization can occur.

Thanks to the progress of experimental techniques, which can nowadays study the coherent dynamics of many-particle quantum systems for long times~\cite{Houck_Nat,Bloch_RMP08,Bloch_Nat}, it has become natural to study ergodicity and thermalization in Hamiltonian quantum systems (see Ref.~\cite{Polkovnikov_RMP11} for a review), a problem dating back to Von Neumann~\cite{vonNeumannTranslation}. The natural tools to study these problems are those developed to analyze the chaotic properties of quantum systems (see Refs.~\cite{Casati:book,Haake:book,Houches} for a review). The dynamics of states in the Hilbert space is linear and therefore cannot be chaotic; chaos can only emerge in the properties of the observables. For instance, the exponential deviations characteristic of chaotic trajectories can be studied through the overlap of the time evolution of the same initial quantum state with two slightly different Hamiltonians (the Loschmidt echo)~\cite{Peres}. Systems whose classical counterpart is chaotic show a Hamiltonian looking like a random matrix and this can be probed from the properties of the level spacing distribution, which is Poisson like for integrable-like systems and Wigner-Dyson for fully ergodic ones~\cite{Bohigas_PRL84,Berry_PRS76}. This analysis has become a probe for chaoticity also in quantum systems without a classical counterpart~\cite{Poilblanc_EPL93}, especially in connection with the recent developments on many body localization~\cite{Basko_Ann06,oganesyan2007localization}. 


A very interesting question addressing the difference between classical and quantum systems is whether quantumness  can modify the chaotic properties of a physical system. A 
remarkable example is the quantum kicked rotor~\cite{Boris:rotor,chirikov,IZRAILEV1990299}.
As discussed in detail in Section~\ref{sec_sr}, the quantum dynamics of this non-linear Hamiltonian model can differ considerably from the corresponding classical one, in terms of ergodicity and energy absorption. 
In the classical case the system behaves ergodically for kicks' amplitudes larger than a critical value: 
the dynamics explores all the available phase space and the energy steadily increases linearly in time without a bound (dynamical delocalization). 
Imposing quantization of conjugated variables, one sees that the energy increases until a certain point and then fluctuates around a finite value (quantum dynamical localization). 
Therefore, quantum interference makes the dynamics of the kicked rotor more regular. This phenomenon is intimately connected to Anderson localization:
quantum interference and chaotic dynamics make the system localized in momentum space (and in energy) in a way similar to one-dimensional Anderson localization in real space~\cite{PhysRev.109.1492}.
The connection between the two phenomena has be discussed in Refs.~\cite{PhysRevLett.49.509, PhysRevA.29.1639}. Dynamical localization in the quantum kicked rotor and in other small chaotic quantum system has also been experimentally observed~\cite{PhysRevLett.61.2011,PhysRevLett.63.364,PhysRevLett.73.2974,PhysRevLett.75.4598,PhysRevLett.117.144104}

The pioneering studies on the quantum kicked rotor done in the seventies and eighties are at the roots of the research field of periodically-driven quantum many body systems. 
Indeed, 
the relation between quantum chaos and ergodicity from one side and energy absorption from the other in this class of systems has recently attracted a lot of interest. This is a very important point for experiments, because periodically driven systems allow to simulate quantum many-body Hamiltonians of physical interest and the dynamics must be stable and non-thermalizing for long times in order to see phenomena like quantum phase transitions and topological effects (see Ref.~\cite{Eckardt_RMP_17} for a review). As in the autonomous case, integrability plays here an important role. General many-body driven quantum systems have been found to attain an asymptotic periodic steady regime described by the so-called Floquet diagonal ensemble~\cite{Russomanno_PRL12}. In the integrable case the steady regime is non-thermal~\cite{Russomanno_PRL12,Russomanno_JSTAT13} (see also~\cite{Arnab_PRB}) and is described by a peculiar form of generalized-Gibbs-ensemble density matrix~\cite{Lazarides_PRL14,russomanno_JSTAT16}. On the other hand, non-integrable driven quantum systems  "thermalize" at $T=\infty$, i.e. heat up indefinitely, because of the absence of energy conservation. Consistently, the eigenstates of the stroboscopic dynamics (the Floquet states) are random delocalized states, locally equivalent to the $T=\infty$ thermal ensemble~\cite{Lazarides_PRE14,Rigol_PRX14}, and the level spacing distribution of the corresponding eigenvalues (the Floquet quasienergies) is of Wigner-Dyson form~\cite{Rigol_PRX14}. For high values of the driving frequency, driven many-body quantum systems can show a long-lived prethermal metastable regime described by the Magnus expansion~\cite{Dalessio_AP13}, which has been shown to be valid only for a finite time~\cite{mori_15,mori_AnnP}. In some systems there is a crossover between thermalizing and integrable-like behaviour for finite size, though it is believed that the dynamics is always eventually thermalizing in the thermodynamic limit~\cite{Rigol_PRX14}, possibly after a prethermalization regime~\cite{Knappa_preprint15,nature_knap}. In other cases, a transition between a regular and an ergodic dynamics persists also in the thermodynamic limit~\cite{Russomanno_EPL15,Buco_arXiv15,Emanuele_2014:preprint}. 
Very peculiar is the case of disordered periodically driven quantum systems, where the transition between a many-body localized regime with dynamical localization and an ergodic thermalizing behaviour is clearly seen~\cite{Ponte_PRL15,Ponte_AP15}. This transition has been experimentally observed~\cite{schn_MBL_per} and its existence has been put in connection with the absence of a mobility edge in the undriven many-body localized model~\cite{Lazarides_PRL15}. In the many-body localized systems, dynamical localization is induced by quantum interference and the {\em disorder imposed externally} on the system. An extremely interesting question is how dynamical localization in clean many-body driven systems~\cite{Russomanno_EPL15,Buco_arXiv15,Emanuele_2014:preprint} is generated by the interplay between quantum mechanics and the {\em disorder spontaneously generated} by the deterministic chaotic dynamics.


In this work we address this question considering a generalization of the quantum kicked rotor to the many body case. Specifically, we study the dynamics of many quantum kicked rotors non-linearly coupled through the kicking. 
Until now only the case of two rotors~\cite{2rot_1, PhysRevLett.110.190401, 2rot_3,2rot_2,PhysRevLett.61.659, PhysRevB.95.064303} and the interacting linear case~\cite{PhysRevB.94.085120} has been considered in literature and a clear picture of the effect of quantum mechanics on the dynamics of the general nonlinear case is missing.
Our goal is to to consider the case of a generic number $N$ of coupled rotors, considering also the thermodynamic limit $N\to\infty$. Our first result is to establish a connection between a chain of $N$ interacting rotors  and a $N$-dimensional disordered lattice exhibiting Anderson localization, extending the results found with $N=1$~\cite{PhysRevLett.49.509, PhysRevA.29.1639} to a generic $N$.
This implies that although the classical system always shows unbounded energy growth, the quantum system can undergo a localization/delocalization transition: also in the many-body case quantum mechanics deeply changes the ergodic properties of the system.
The connection is first explored analytically and afterwards numerically in the cases $N=2$ and $N=3$ using exact diagonalization and a time-evolving-block-decimation algorithm. 
For $N=2$, we find that the energy initially increases in time, as it was previously found~\cite{2rot_1,PhysRevLett.61.659} eventually stopping to a finite asymptotic value exponentially large in the kicking strength, therefore  exhibiting dynamical localization. This result is in agreement with the results for the two-dimensional  Anderson model~\cite{Licciardello}. 
For $N>2$ it is known that a disordered lattice undergoes a transition from Anderson localization~\cite{Licciardello}: we numerically observe this in the case of $N=3$. These results for the rotors are pictorially represented in FIG.~\ref{fig:intro}. 



Finally we move to the large $N$ limit. 
We first study an $\infty$-dimensional Anderson model: using the scaling theory of localization by Abrahams {\em et. al.}~\cite{Licciardello} we show that this model always exhibits delocalization. Therefore we expect, due to the mapping introduced above, that also the many kicked rotors model is always dynamically delocalized in the thermodynamic limit: the energy per site always increases without a bound. 

We perform a numerical study of this limit with a mean-field approximation which is exact when the coordination number of the system goes to infinity: we focus on a specific case in which this fact occurs, the one of infinite-range interactions in the thermodynamic limit. 
With the mean-field approach we can use an effective single rotor Hamiltonian to infer the dynamics of the long range interacting system. 
We remarkably find that the system is not localized: the momentum distribution spreads in time and the kinetic energy grows. 
This growth is described by an anomalous diffusion, namely the energy increases like $t^\alpha$ with $\alpha<1$.
For high values of kicking amplitude and interaction we find that subdiffusion tends to become diffusion: $\alpha \rightarrow 1$. The subdiffusion we observe is a genuinely quantum phenomenon: for the same parameters the classical counterpart of the system is ergodic and its energy grows  linearly in time (diffusive behaviour).

The peculiarity of the effective single rotor model is that the kick amplitude evolves in time: it 
is modulated by a mean field parameter which is computed at each time step and depends from the evolution of the system itself. 
The breaking of the dynamical localization in a single quantum kicked rotor via a modulation of the kick amplitude has already been considered. 
Examples are a modulation via $d-1$ incommensurate frequencies~\cite{PhysRevLett.115.240603,PhysRevLett.62.345, Yamada_PRE, PhysRevE.79.036206, PhysRevA.80.043626} which induces Anderson localization/delocalization transition and a kick with modulated amplitude which undergoes decoherence and quantum-to-classical transition~\cite{PhysRevLett.80.4111, PhysRevLett.98.260401, PhysRevA.77.062113}. In all the cases, the properties of the modulation are crucial in determining
the response of the system, especially if the modulation is
noisy~\cite{PhysRevLett.87.074102, PhysRevE.61.7223, PhysRevLett.81.1203,  PhysRevE.70.036217, 1464-4266-2-5-307, PhysRevA.65.032115}.
In our case the modulation does not come from an external signal but is self-consistently determined.  
Moreover, the mean field parameter introduces a nonlinearity in the effective Hamiltonian which plays a crucial role in destroying the dynamical localization of the single rotor. 
The non-linearity induces  a self-reorganization during the system evolution giving rise to the anomalous diffusion of the kinetic energy.
Nonlinearities have already been considered in the kicked
rotor~\cite{refId0, d4freq, PhysRevLett.100.094101} and related disordered lattices~\cite {PhysRevLett.70.1787, Nguyen2011,PhysRevLett.112.170603, Qin2017} and
they are indeed found to turn the dynamical or Anderson
localization into a subdiffusive spreading of the wave function.

The work is organized as follows. In Section~\ref{sec_sr} we introduce the interacting kicked rotors model we discuss in this work. We also briefly review the single kicked rotor, focusing on the different behaviors that the classical and quantum versions of this model manifest. 
In Section~\ref{mapping:sec} 
we discuss the analytical mapping of the $N$-rotors model on an $N$-dimensional Anderson model and numerically verify it in $N=2$ and $N=3$ rotors cases. A comparison with the results for the classical case is reported in Appendix~\ref{classical:sec}.
Section~\ref{thermodyno:eqn} discusses the behaviour of the coupled kicked rotors in the thermodynamic limit. 
We predict that in this case there is always dynamical delocalization: we show this in Subsection~\ref{noloc_inf:sec} where we use the mapping introduced in Section~\ref{mapping:sec} and demonstrate the absence of Anderson localization for an $\infty$-dimensional disordered lattice.
Section~\ref{mean_field1:sec} contains the numerical study of the $N\rightarrow \infty$ limit of the fully connected model: we define the effective mean field model (for a demonstration of its exactness when $N\rightarrow \infty$ see Appendix~\ref{mf_calc}) and describe its dynamics. We study the dynamics of the effective model by looking at the kinetic energy growth and at the properties of the time dependent mean field parameter.
In the Conclusions we summarize the results henceforth presented, discussing the outlook and the implications coming from this work.

%

\section{kicked rotors models}\label{sec_sr}

The kicked rotor (KR) is a paradigmatic model both in classical and quantum mechanics, widely studied from the appearance of the first works~\cite{Chirikov1979263, chirikov,PhysRevLett.49.509, PhysRevA.29.1639} 
. For a review one could see Ref.~\cite{Delande_notes}.
Here we study a many-body generalization of this model, whose adimensional Hamiltonian is 
\begin{equation}\label{H_a:eqn}
\hat{H}(t) = \frac{1}{2}\sum_{i=1}^N \hat{p}_i^2+V(\hat{\boldsymbol\theta})\sum_{n=-\infty}^{+\infty}\delta (t-n)\,;
\end{equation}
with
\begin{equation} \label{H_sr:eqn}
 V(\hat{\boldsymbol\theta}) = 
  K\left( \sum_{i=1}^N \cos \hat{\theta}_i 
 -\frac{1}{2} \sum_{i\neq j}\epsilon_{ij} \cos (\hat{\theta}_i-\hat{\theta}_{j})\right)\,.
\end{equation}
In this work we specifically address two cases: the one with nearest neighbour interactions 
where $\epsilon_{ij}=\epsilon\,\delta_{i,j-1}$ and that  of infinite-range interactions 
$\epsilon_{ij}=\frac{\epsilon}{N-1}$. In the latter case the mean-field approximation is exact in the thermodynamic limit.
%
%
%
%
Notice the commutation rules 
\begin{equation}\label{com:eqn}
[\hat\theta_i, \hat p_j]=i\kbar\delta_{i,\,j}.
\end{equation}
where the effective Planck's constant $\kbar=\hbar\, T/I$ is directly proportional to $\hbar$ and to the physical kicking period $T$ and inversely proportional to the momentum of inertia $I$ of the rotors~\cite{Boris:rotor}.
This adimensional constant is obtained by expressing the Hamiltonian Eq.~\eqref{H_a:eqn} in units of $I/T^2$, 
defining the following adimensional quantities: $t'= t/T$, $K'= T\,K/I$, $\hat p' = \hat p \, T/I$. After this rescaling, the kicking period is 1, as can be seen in Eq.~\eqref{H_a:eqn}. We will henceforth be interested in the stroboscopic evolution of the system at each period of time: we consider the state of the system only at discrete times $t_n\equiv n$.
 
The momentum operators
$\hat{p}_i$ have discrete eigenvalues $\kbar m_i$ ($m_i\in\mathbb{Z}$) as a result of the corresponding angle operator being periodic $\hat{\theta_i}=\hat{\theta_i}+2\pi$ and the wave-function in the angle representation being single-valued (see for instance Ref.~\cite{Picasso:book}). A possible basis of the Hilbert space is therefore easily constructed from  tensor products of local momentum eigenstates
$\{\ket{m_1,\ldots,m_N}\}_{m_1,\ldots,m_N\in \mathbb{Z}}$. We will write this basis also in the form $\{\ket{\bf m}\}_{{\bf m}\in \mathbb{Z}^N}$ where we have defined the vector ${\bf m}\equiv \left(m_1,\ldots,m_N\right)$.

Before moving to the analysis of the many-rotors models, let us review what is known about the single KR ($N=1$).
%
%
Classically, this model can either show energy localization or unbounded energy
growth depending on the value of $K$.
This can be seen by studying the stroboscopic kinetic energy of the system evaluated immediately before the $n$-th kick which is $E(n)=\overbar{p^{\,2}(n)}/2$ (the average $\overline{(\cdot)}$ is taken over an ensemble of randomly chosen initial conditions). 
The energy will not increase in time if $K<K_c=0.971635$
(classical dynamical localization) due to the presence of stable KAM trajectories separated by chaotic regions~\cite{doi:10.1063/1.524170}. 
These stable trajectories disappear for $K\gtrsim K_c$ and the dynamics becomes fully chaotic; 
as a consequence $E(n)$ starts growing linearly in time with a coefficient $D_{KR}\simeq K^2/4$. 
In this regime the system is ergodic: nearby trajectories 
separate exponentially and explore the entire phase space for generic initial conditions. As a consequence, there is diffusion in the momentum space, as it can be seen looking at the momentum variance $\sigma_p^2(n)\equiv\overline{{p^2}(n)}=2E(n)$~\cite{Note_sig} which increases linearly with $n$. Since this object coincides with the kinetic energy up to a factor, from a classical point of view ergodicity implies energy delocalization.

The quantum counterpart of this model (quantum kicked rotor -- QKR) is obtained by 
imposing the commutation rules Eq.~\eqref{com:eqn} to the case $N=1$. 
Quantum mechanics dramatically changes the behavior of this model killing ergodicity and constrains the energy dynamics so that the system behaves as an integrable one.
Indeed we pass from the unbounded steady heating of the classical system to dynamical localization for all values of $K$ exhibited by its quantum counterpart.
The kinetic energy, after a linear growth for a time $n^*$~\cite{CHIRIKOV198877, PhysRevA.29.1639}, reaches an asymptotic condition and fluctuates around a finite value.~\cite{NOte10}
Dynamical localization has been experimentally observed for the first time with a cloud of ultracold atoms moving in a pulsed, one dimensional periodic optical lattice~\cite{PhysRevLett.75.4598}.
%

The dynamical localization in the QKR can be better understood with the mapping introduced in \cite{PhysRevLett.49.509, PhysRevA.29.1639}, which connects this model  to the time-independent Hamiltonian of a single particle hopping on a disordered one dimensional lattice. This last model is known to show Anderson localization~\cite{PhysRev.109.1492}: the eigenfunctions at energy $\epsilon$ are localized in space, $\psi_\epsilon(x)\sim\exp(-x/\xi)$, where $\xi$ is the localization length.
Such construction will be generalized to the many-rotors models defined in  Eq.~\eqref{H_sr:eqn} in the next section making it possible to interpret dynamical localization/delocalization in these models in terms of Anderson localization of a particle hopping over an $N$-dimensional lattice. 
\section{Floquet states and mapping to Anderson localization} \label{mapping:sec}
Our first step in the analysis of the behaviour of coupled quantum kicked rotors is to
develop a mapping of a model of $N$ kicked rotors to a single particle hopping in a $N$-dimensional disordered lattice model (Subsection~\ref{construction:subsec}). Using Floquet states in a way similar to what Refs.~\cite{PhysRevLett.49.509, PhysRevA.29.1639} do for a single rotor, we will show that the hopping in the lattice model is short-ranged for all the cases we are interested in allowing us to apply the existing knowledge on the localization/delocalization transition. We show that localization/delocalization in the lattice model precisely corresponds to dynamical localization/delocalization in the rotors model (Subsection~\ref{dynamo:subsec}). 
We can therefore make the following  predictions: 
for $N\leq2$,  the lattice model is always Anderson localized -- and so should be the rotors dynamics in the energy space. For $N>2$ the lattice model undergoes a transition from localization  to delocalization as the hopping strength is increased~\cite{Licciardello} implying a dynamical localization/delocalization transition for the rotors. For $N=2$, we expect the asymptotic energy to be exponentially large in the kicking strength. In Subsection~\ref{numerical:subsec} we numerically verify our predictions for the kicked rotors  in the cases $N=2$ and $N=3$. We do this by studying the energy dynamics, the inverse participation ratio of the Floquet states and the level spacing distribution.
%
\subsection{Localization of the Floquet states} \label{construction:subsec}
%

In order to present the mapping of our model to an Anderson one let us start by studying the properties of the time-evolution operator over one period. We consider the evolution from the instant immediately before the $n$-th kick to the instant immediately before the $n+1$-th. The desired time evolution operator is therefore
\begin{equation} \label{lr_U:eqn}
  \hat{U} = \exp\left(-\frac{i}{2\kbar}\sum_{i=1}^N \hat{p}_i^2\right)\exp\left(-\frac{i}{\kbar}V(\hat{\boldsymbol\theta})\right)\,,
\end{equation}
where $\kbar$ is the effective Planck's constant introduced in Section~\ref{sec_sr}. 
Let us now focus on the properties of the eigenstates of this evolution operator, the so-called Floquet states $\ket{\phi_\alpha}$~\cite{Samba,Shirley_PR65}.  In
other words%
\begin{equation} \label{eigenstates:eqn}
  \hat{U}\ket{\phi_\alpha}=\nep^{-i\mu_\alpha}\ket{\phi_\alpha}\,,
\end{equation}
where $\mu_\alpha$ are the Floquet quasienergies. The Floquet states $\ket{\phi_\alpha}$ are eigenstates of the stroboscopic dynamics which therefore are left invariant up to a phase factor by the action of $\hat{U}$. 
 Let us now define $\hat{H}_0\equiv\frac{1}{2}\sum_{i=1}^N\hat{p}_i^2$ (see Eq.~\eqref{H_a:eqn}) and apply the unitary transformation $\ket{\widetilde{\phi}_\alpha}=\nep^{i\hat{H}_0/(2\kbar)}\ket{{\phi}_\alpha}$. We can apply this transformation without altering the localization structure of the Floquet state in the momentum basis, because the operator $\hat{H}_0$ is diagonal in this basis. 
After the transformation, we can rewrite the eigenvalue equation as a pair of equations~\cite{NOte5} 
%
\begin{equation}
   \nep^{\mp i\hat{H}_0/(2\kbar)}\exp\left(\mp\frac{i}{\kbar}{V}(\hat{\boldsymbol \theta})\right)\nep^{\mp i\hat{H}_0/(2\kbar)}\ket{\widetilde{\phi}_\alpha} = \nep^{\mp i\mu_\alpha}\ket{\widetilde\phi_\alpha}\,.
\end{equation}
%
%
%
Using the resolution of the identity
\begin{equation} \label{resolution:eqn}
  \boldsymbol{1}=\sum_{\bf m}\ket{\bf m}\bra{\bf m}
\end{equation}
in terms of the momentum eigenstates
 and performing some simple formal manipulations, we can finally rewrite the Floquet eigenvalue equation as
\begin{equation} \label{eigenstat:eqn}
  \sum_{{\bf m}'\neq {\bf m}}W_{{\bf m}\,{\bf m}'}\bra{{\bf m}'}\widetilde\phi_\alpha\rangle+\epsilon({\bf m})\bra{{\bf m}}\widetilde\phi_\alpha\rangle=  2 \cos(\mu_\alpha)\bra{{\bf m}}\widetilde\phi_\alpha\rangle\,,
\end{equation}
with
\begin{align} \label{hoppo:eqn}
 & W_{{\bf m}\,{\bf m}'}=\\ \nonumber
 & 2\mathrm{Re}\left[\nep^{-i(\varphi({\bf m}')+\varphi({\bf m}))}\int\frac{\ud^N\boldsymbol{\theta}}{(2\pi)^N}\nep^{-{i}{V}({\boldsymbol \theta})/{\kbar}}\nep^{-i({\bf m}'-{\bf m})\cdot\boldsymbol\theta}\right]\,, 
\end{align}
%
and
\begin{align}
  \epsilon({\bf m}) &= 2\mathrm{Re}\left[\nep^{-2i\varphi({\bf m})}\int\frac{\ud^N\boldsymbol{\theta}}{(2\pi)^N}\nep^{-{i}{V}({\boldsymbol \theta})/{\kbar}}\right]; \\
  \quad \varphi({\bf m}) &= \frac{\kbar}{4}\sum_{i=1}^Nm_i^2\,.
\end{align}
Equation (\ref{eigenstat:eqn}) can be seen as the static Schr\"odinger equation of a particle hopping in an $N$-dimensional potential $\varepsilon({\bf m})=W_{{\bf m}\,{\bf m}}$. This potential behaves as a true disorder for localization purposes in a one-dimensional next-nearest-neighbor tight-binding model~\cite{NOte3}. This is true for all values of $\kbar$ but the integer multiples of $4\pi$: In this case the potential $\varepsilon({\bf m})$ is constant and cannot induce any localization. 
The mapping we propose is different from the one introduced in Refs.~\cite{PhysRevLett.49.509, PhysRevA.29.1639} for the single kicked rotor: In our case the hopping does not show unphysical divergences which instead occur in Refs.~\cite{PhysRevLett.49.509, PhysRevA.29.1639} due to the small convergence radius of the Fourier series of the tangent. Our wave-function $\bra{{\bf m}'}\widetilde\phi_\alpha\rangle$ is normalized { by construction} and there is no risk for spurious unphysical divergences in the hopping because the integrand in Eq.~\eqref{hoppo:eqn} is always bounded in modulus.
In the single-rotor ($N=1$) case, the hopping is
\begin{equation}\label{1d_hop:eqn}
  W_{{m}\,{m}'} = 2J_{m'-m}\left(\frac{K}{\kbar}\right)\mathrm{Re}[i^{m'-m}\nep^{i(\varphi(m)+\varphi(m'))}]
\end{equation}
($J_{m'-m}$ is the Bessel function of order $m'-m$). The modulus of this expression always decays faster than exponentially~\cite{Stegun} with $m'-m$ and never shows unphysical divergences.  We have therefore 
a one-dimensional Anderson model which is always localized~\cite{PhysRev.109.1492}. 

Now we go through  the analysis of the hopping coefficients $W_{{\bf m}\,{\bf m}'}$ for $N>1$.
We observe that, although they depend on ${\bf m}$ and ${\bf m}'$ separately, they are symmetric under the parity transformation $({\bf m},{\bf m}')\rightarrow (-{\bf m},-{\bf m}')$ and  permutations of the Cartesian components of ${\bf m}-{\bf m}'$ (namely, ${\bf m}$ and ${\bf m}'$ undergo the same permutation of the components).
Also, the $ W_{{\bf m}\,{\bf m}'}$ depend on the direction given by the vector ${\bf m}-{\bf m}'$.
%

\begin{figure}
    \hspace{-0.45cm}
    \resizebox{\columnwidth}{!}{\includegraphics{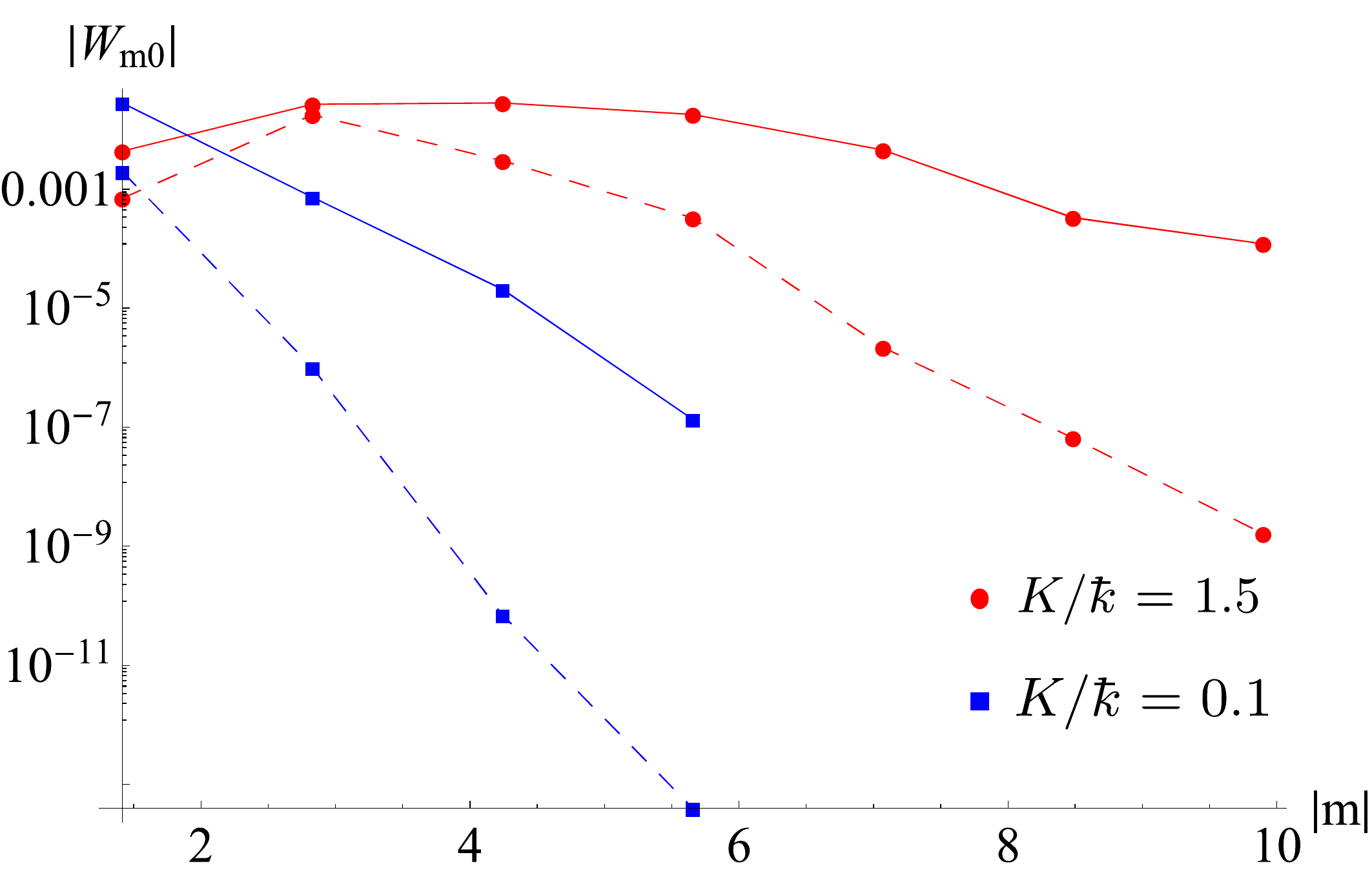}}
  \caption{\label{fig:W_vs_m}The modulus of  $W_{{\bf m}\,\boldsymbol{0}}$ is plotted as a function of $|{\bf m}|$ for $K/\kbar=0.1$ (blue squares) and $K/\kbar=1.5$ (red circles) for $N=3$. The continuous and dashed lines correspond to two orthogonal directions in ${\bf m}$ space. We see the slope of the exponential decaying which decreases as $K/\kbar$ is increased; an analogous behavior is found with $N=2$.} 
  \end{figure}

In order to apply existing results on the Anderson model to the case $N>1$ we need to verify that the hopping is short ranged, as it is for $N=1$. The exact analytical expression for $W_{{\bf m}\,{\bf m}'}$ cannot be established, nevertheless we can infer information on how this hopping coefficient decays with $|{\bf m}-{\bf m}'|$ by observing that $W_{{\bf m}\,{\bf m}'}$ is the linear combination of real and imaginary part of the Fourier transform of $f(\bf{\theta})=\nep^{-{i}{V}({\boldsymbol \theta})/{\kbar}}$. Since the function $f(\bf{\theta})$ is $\mathcal{C}(\infty)$ and it is $2\pi$-periodic in all the $\theta_j$,  its Fourier components $\hat f(\bf{m}'-\bf{m})$ decay exponentially fast with $|\bf{m}'-\bf{m}|$. It means that the hopping $W_{{\bf m}\,{\bf m}'}$ is short ranged and therefore the results concerning the Anderson model can always be applied to the effective hopping model Eq.~\eqref{eigenstat:eqn}. 
In the following we numerically show the decaying properties of the hopping coefficients for $N=2$ and $N=3$. In FIG.~\ref{fig:W_vs_m} we set ${\bf m}'=0$  and plot the behavior of the hopping coefficients $|W_{{\bf m}\,\boldsymbol{0}}|$ as a function of the distance $|{\bf m}|$ for $K/\kbar=0.1$ and $K/\kbar=1.5$; ${\bf m}$ are taken along two orthogonal directions (dashed and continuous lines in FIG.~\ref{fig:W_vs_m}). We find that
 the exponential decay is clearly seen and is smaller for increasing values of $K/\kbar$. This behavior is the same along the two directions even if the values of $|W_{{\bf m}\,\boldsymbol{0}}|$ are different due to the space anisotropy.

In order to quantify the strength of the hopping we define two quantities: the first is the hopping integral $\Sigma$
\begin{equation}\label{int_range}
\Sigma = \sum_{{\bf m}\in\mathbb{Z}^N}|W_{{\bf m}\,\,\boldsymbol{0}}|.
\end{equation}
For a short ranged lattice we expect this quantity to be finite at fixed $K/\kbar$, while it diverges if the hopping is long ranged. The second is the hopping range, defined as
\begin{equation}\label{hop_range}
\rho = \frac{\sum_{{\bf m}\in\mathbb{Z}^N}|W_{{\bf m}\,\,\boldsymbol{0}}|\,|{\bf m}|}{\Sigma}.
\end{equation}
We say that the hopping strength of the lattice model increases when the hopping integral and the hopping range are increased.
We first verify that the hopping in the lattice Eq.~\eqref{eigenstat:eqn} is short ranged and therefore well defined. In addition the hopping strength, estimated through the hopping integral and range (see Equations~\eqref{int_range} and~\eqref{hop_range}) is found to be monotonously increasing as a function of $K/\kbar$.
%
%
%


Let us discuss the numerical computation leading to these results (in the rest of the discussion $\kbar=400$ and 
$\epsilon = -2$ for definiteness). 

%
%
\begin{figure*}
\centering
\includegraphics[width=12cm]{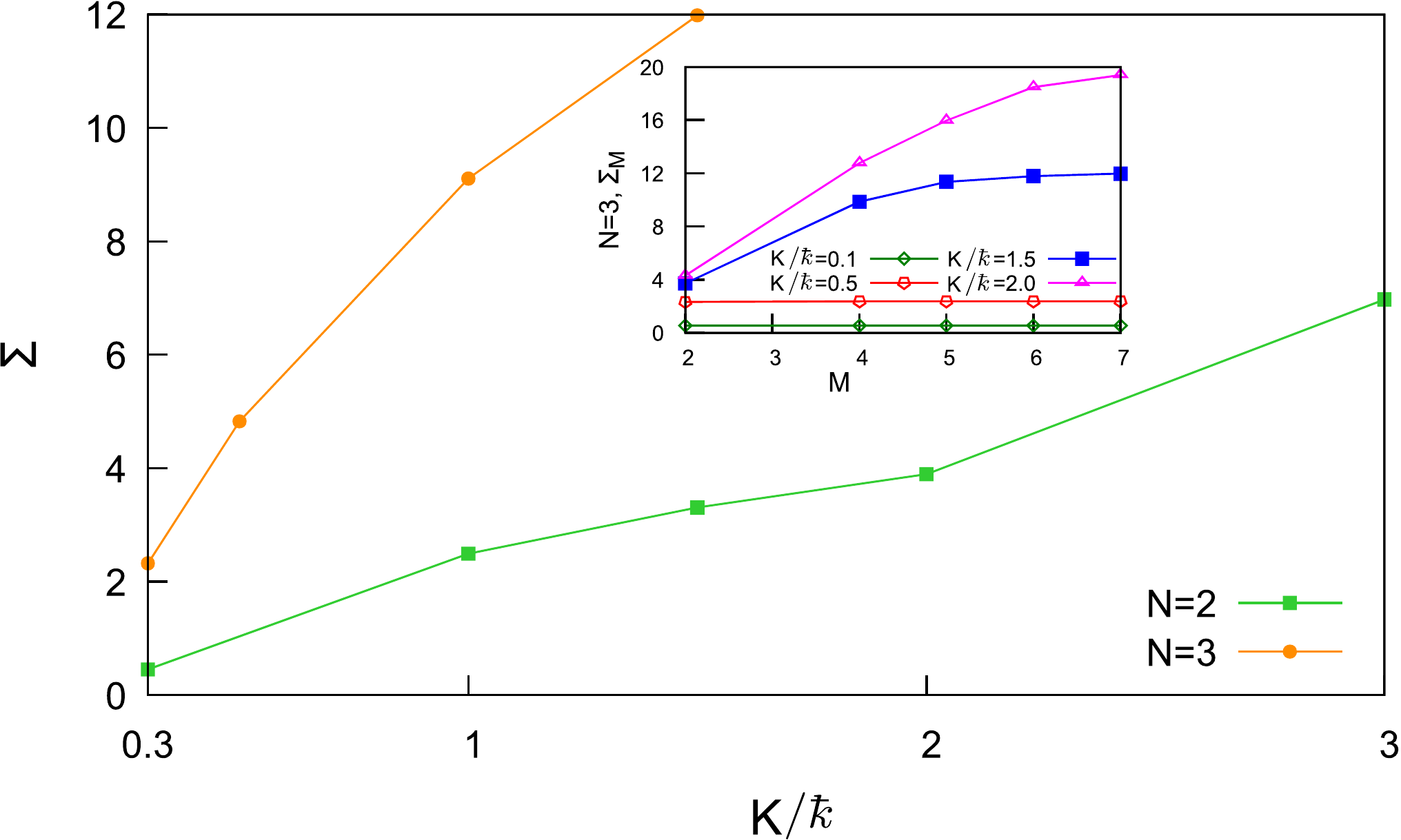}
\caption{\label{fig:int_range} 
The hopping integral $\Sigma$ vs. $K/\kbar$ is plotted for $N=2$ (green squares) and $N=3$ (yellow circles). 
At fixed $N$ it exists a maximum value of $M$ for which the convergence of the integrals $W_{\bf m}$ can be numerically achieved: this imposes a limit to the maximum value of $K/\kbar$ for which $\Sigma$ can be computed. 
This is shown in the inset for $N=3$; since for $N=3$ the maximum value is $M=7$, then the convergence of $\rho_M$ can be observed only up to $K/\kbar=2.0$. For the $N=2$ case the maximum value is $M=20$. The resulting hopping integral $\Sigma$ is a monotonically increasing function of $K/\kbar$ as shown in the main figure. }
\end{figure*}

\begin{figure*}
\centering
\includegraphics[width=12cm]{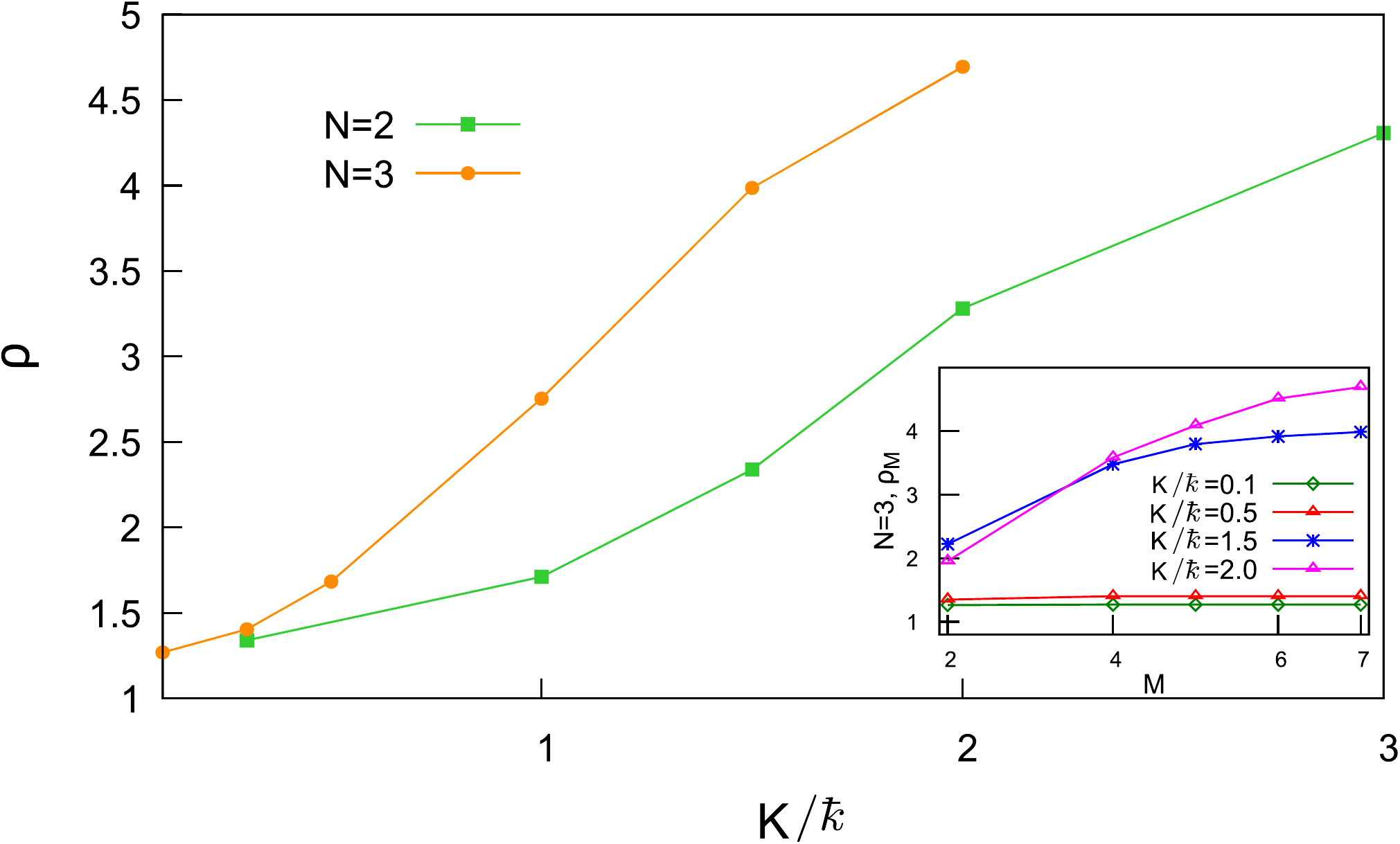}
\caption{\label{fig:hop_range} The hopping range $\rho$ vs. $K/\kbar$ for $N=2$ (green squares) and $N=3$ (yellow circles): it grows as $K/\kbar$ is increased. 
As for the computation of $\Sigma$, in the the numerical computation of $W_{\bf m}$ we have a maximum $M=7$ for $N=3$ implying that the convergence of $\rho$ is observed only for $K/\kbar \leq 2.0$ (see inset). For the case $N=2$ the interaction range is plotted up to $K/\kbar=3.0$ where convergence is observed with $M=18$. Notice that also in this case $\rho$ is an increasing function of $K/\kbar$.}
\end{figure*}

The exponential decay of the hopping strength makes the hopping integral defined in Eq.~\eqref{int_range} finite: we compute it by taking the asymptotic value $\Sigma_\infty\equiv\Sigma$ of the series
\begin{equation}\label{int_range}
\Sigma_M =\sum_{{\bf m}\in\mathcal{C}(M)}|W_{{\bf m}\,\boldsymbol{0}}|\,, \ \ M\in\mathbb{N} 
\end{equation}
where $\mathcal{C}(M)$ is the $N-$dimensional cube with edge length $2M$ centered in ${\bf 0}$. The hopping integral  is plotted in FIG. \ref{fig:int_range} as a function of $K$ for $N=2$ and $N=3$. The inset shows some examples of convergence of $\Sigma_M$ for increasing values of $M$, for some values of $K/\kbar$ and $N=3$. In the case $N=2$ the behavior is the same, except that higher values of $M$ have to be considered to achieve the convergence (the limitation on the value of $M$ has computational reasons due to the possibility to compute $W_{{\bf m}}$ up to a certain ${\bf m}$ with a maximum error $\sim10\%$).
In a similar way we compute the hopping range defined in Eq.~\eqref{hop_range} to find that it is finite: we consider the series
\begin{equation}\label{int_range}
\rho_M = \frac{\sum_{{\bf m}\in\mathcal{C}(M)}|W_{{\bf m}\,\boldsymbol{0}}|\,|{\bf m}|}{\Sigma_M}\,, \ \ M\in\mathbb{N} 
\end{equation}
and check its convergence as $M$ is increased. In FIG. \ref{fig:hop_range} $\rho$ is plotted as a function of $K/\kbar$ for $N=2$ and $N=3$. The inset shows also in this case the convergence of $\rho$ as a function of $M$ for some values of $K/\kbar$. 

In conclusion, we have provided an analytical argument to state that the lattice model is short ranged. 
Moreover, we have numerically checked this property in the interval of $K/\kbar$ we have access to, finding also that the hopping strength (i.e. both $\Sigma$ and $\rho$) increases with $K/\kbar$. 
Therefore we can apply the general theory on Anderson localization~\cite{PhysRev.109.1492,Licciardello} and we predict that for $N=2$ our model will display localization with a localization length exponentially large in $K/\kbar$, while it will undergo a localization/delocalization transition at some value of $(K/\kbar)_c(N)$ when $N\geq 3$.
In the next subsection we are going to show how the localization properties of the Floquet states in the momentum space do indeed reflect on the dynamical localization of the energy.

\subsection{Dynamical localization and Floquet states} \label{dynamo:subsec}
In order to understand the connection between localization in momentum space and dynamical localization let us express the energy in terms of the Floquet states. We start the dynamics from the ground state of the kinetic energy operator, the state with all vanishing local momenta $\ket{\Psi_0}\equiv\ket{\boldsymbol0}$; we can therefore expand  the time-evolved state immediately before the $n$-th kick -- $\ket{\Psi(n)}\equiv\hat{U}^n\ket{\Psi_0}$ --in the basis of the Floquet states as~\cite{NOte2}
\begin{equation} \label{evoluzione:eqn}
  \ket{\Psi(n)} = \sum_{\alpha=-\infty}^\infty  \nep^{-i\mu_\alpha n}\ket{\phi_\alpha} \bra{\phi_\alpha}\left.\boldsymbol{0}\right\rangle
\end{equation}
%
Using this expansion, we can express the energy per site immediately before the $n$-th kick -- $E(n)\equiv\bra{\Psi(n)}\hat{H}_0\ket{\Psi(n)}/N$ -- in the form
\begin{equation} \label{Henne:eqn}
  E(n)=\frac{1}{N}\sum_{\alpha,\beta=-\infty}^\infty \left\langle\boldsymbol{0}\right.\ket{\phi_\alpha}\bra{\phi_\beta}\left.\boldsymbol{0}\right\rangle\bra{\phi_\alpha}H_0\ket{\phi_\beta}\nep^{i(\mu_\alpha-\mu_\beta)n}\,.
\end{equation}
The system is dynamically localized if, after a transient, this object fluctuates around a {\em finite} value given by the infinite-time average
\begin{equation} \label{timeav0:eqn}
  E_{\rm av}(\infty)=\lim_{\mathcal{T}\to\infty}E_{\rm av}(\mathcal{T}) = \lim_{\mathcal{T}\to\infty}\frac{1}{\mathcal{T}}\sum_{n=0}^{\mathcal T}E(n).
\end{equation}
Using Eq.~\eqref{Henne:eqn} for $E(n)$, the resolution of the identity Eq.~\eqref{resolution:eqn} and assuming no degeneracies in the Floquet spectrum, we can evaluate this average as
\begin{equation} \label{timeav:eqn}
  E_{\rm av}(\infty)=\frac{\kbar^2}{2N}\sum_{\alpha} |\left\langle\boldsymbol{0}\right.\ket{\phi_\alpha}|^2\sum_{\bf m}|\left\langle{\bf m}\right.\ket{\phi_\alpha}|^2\sum_{j=1}^Nm_j^2\,.
\end{equation}
%
%
If the Floquet states are localized in the momentum basis then the wave-function in this basis will behave as $\left\langle{\bf m}\right.\ket{\phi_\alpha}\simeq\mathcal{N}\nep^{-|\mathbf{m}-\mathbf{m}_\alpha|/\lambda}$ for some $\lambda$ and $\mathbf{m}_\alpha$ ($\mathcal N$ is some normalization factor). Assuming that the localization centres $\mathbf{m}_\alpha$ are uniformly distributed with density $\sigma$ in the $N$-dimensional space, we can give an estimate of the time-averaged energy~\cite{NOte4}
\begin{equation} \label{aver2:eqn}
  E_{\rm av}(\infty)\sim\kbar^2\frac{\sigma(N+1)}{4}\lambda^{2}
\end{equation}
%
%
which is finite if the momentum localization length of the Floquet states $\lambda$ is finite. 
Therefore localization of Floquet states in the momentum basis implies dynamical localization of energy. Therefore the mapping of Subsection~\ref{construction:subsec} makes us predict the existence of a dynamical localization/delocalization transition at some $K_c(N)$ when $N\geq 3$, while the system is always dynamically localized for $N\leq 2$: in the next subsection we are going to numerically verify these predictions for the cases $N=2$ and $N=3$.
\subsection{Numerical results} \label{numerical:subsec}
\subsubsection{Energy dynamics}
For the study of the dynamics of the model Eq.~\eqref{H_sr:eqn} we use two numerical methods: exact diagonalization for $N=2$ and time-evolving block decimation (TEBD) on matrix product states (MPS)~\cite{Daley_JSTAT04,Schollwock_rev} for $N=3$. In both cases we need to truncate the Hilbert space, whose dimension is a countable infinity. We truncate it in the momentum basis: selecting a cutoff $M$ (called "local truncation dimension") we impose that the time-evolving state is a superposition of momentum states $\ket{m_1,\ldots,m_N}$ with $-M\leq m_j\leq M$. We  evolve with the Hamiltonian restricted to this subspace. If the system is dynamically localized, high momentum will never be involved in the dynamics: provided that $M$ is big enough our numerics will correctly describe the dynamics even for long times. On the opposite, if there is dynamical delocalization, our simulations will be meaningful up to a certain time.

\begin{figure}
  \begin{center}
    \begin{tabular}{c}
     \hspace{0cm}\resizebox{85mm}{!}{\includegraphics{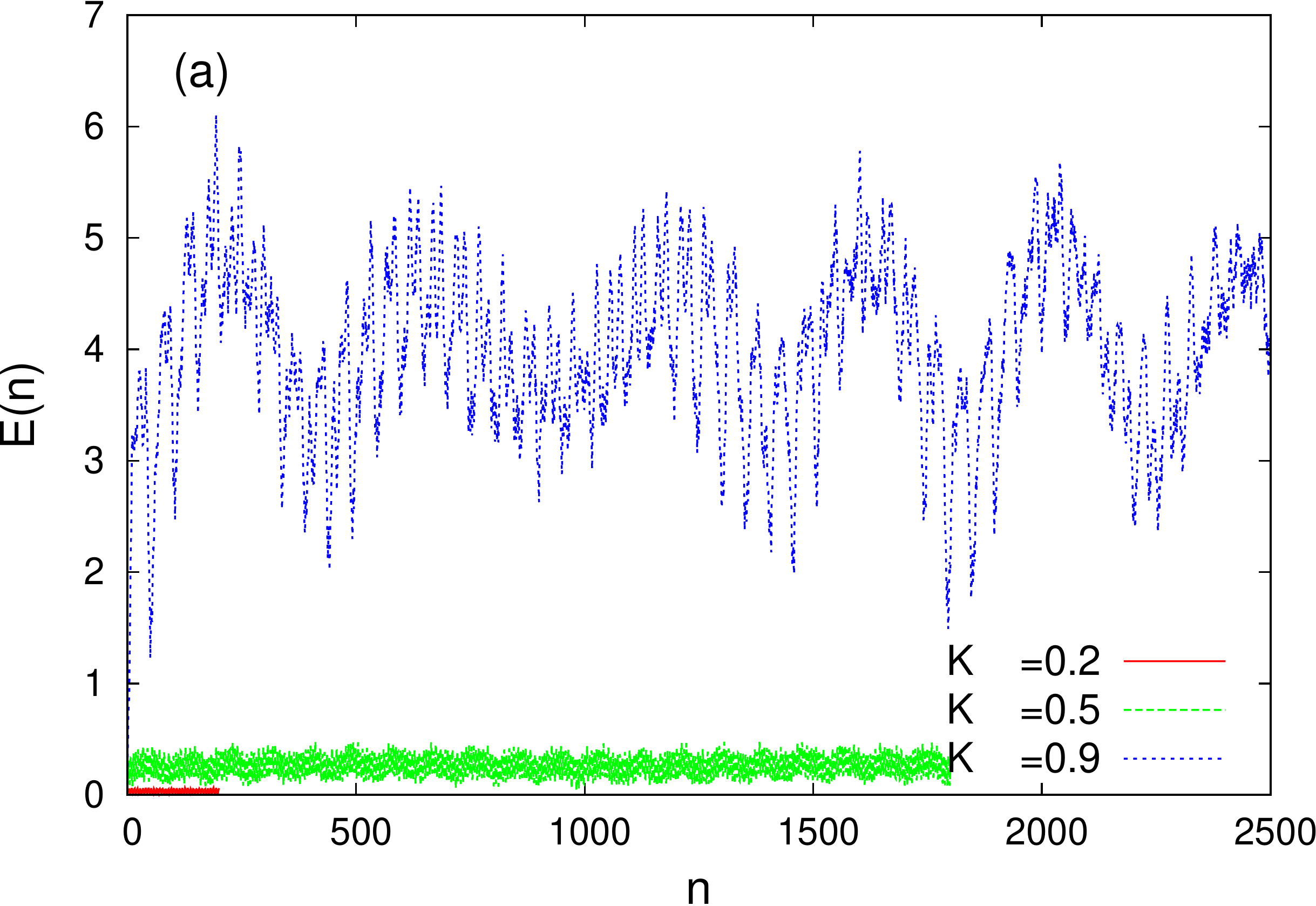}}\\
     \pgfputat{\pgfxy(2.1,2.1)}{\pgfbox[left,top]{\footnotesize $/\kbar$}}
     \pgfputat{\pgfxy(2.0,1.75)}{\pgfbox[left,top]{\footnotesize $/\kbar$}}
     \pgfputat{\pgfxy(1.9,1.44)}{\pgfbox[left,top]{\footnotesize $/\kbar$}}\\
     \hspace{-0.5cm}\resizebox{85mm}{!}{\includegraphics{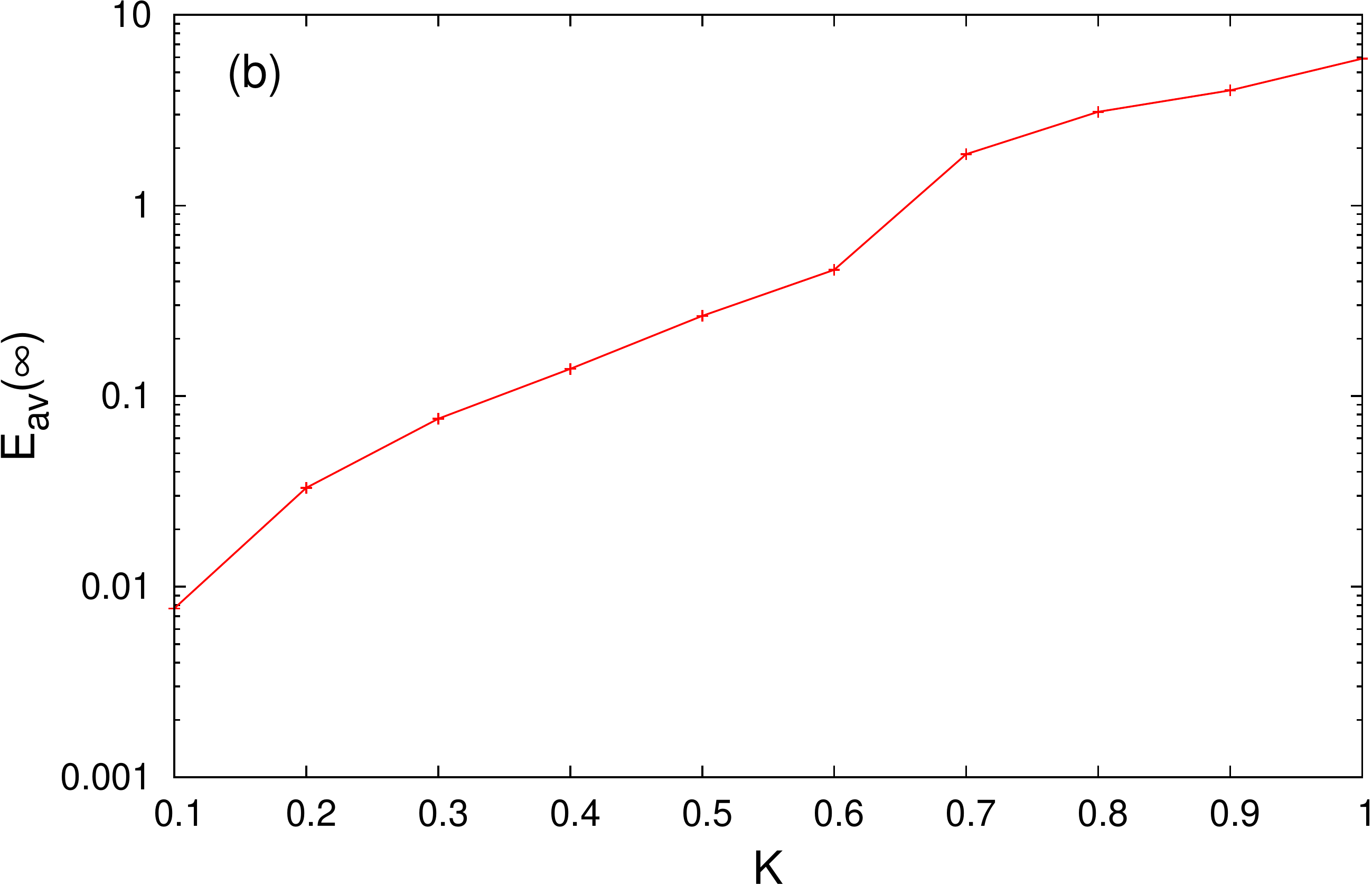}}\\
     \pgfputat{\pgfxy(0.4,0.61)}{\pgfbox[left,top]{\small $/\kbar$}}\\
    \end{tabular}
  \end{center}
  \caption{(a) Energy dynamics for $N=2$ and different values of $K$ obtained with exact diagonalization. For the considered values of $K/\kbar$ we always see dynamical localization. (b) Time-average $E_{\rm av}(\infty)$ versus $K/\kbar$: we see that it increases exponentially. In all the cases we take $M\leq 24$, big enough so that the time traces have converged in $M$. Numerical parameters: $\kbar=400$, $\epsilon =-2$.
  }
  \label{2:fig}
\end{figure}

In FIG.~\ref{2:fig} we report exact diagonalization results for $N=2$. In panel (a) we show some examples of energy evolution: we always observe localization  (we take $M$ big enough so that the energy time-trace is converged). In order to  estimate the infinite-time-averaged energy Eq.~\eqref{timeav0:eqn}, in panel (b) we plot the time-averaged energy ${E}_{\rm av}(\mathcal{T})$ over a time $\mathcal{T}\gg 1$ versus $K/\kbar$.
%
%
Since for large $\mathcal T$ this function tends to converge choosing a  large enough $\mathcal T$ we can extract a good estimate of ${E}_{\rm av}(\infty)$. We see that $E_{\rm av}(\infty)$ exponentially increases with $K$ giving rise to a localization length $\lambda$ exponentially large in $K$ (see Eq.~\eqref{aver2:eqn}); this confirms our predictions in the case $N=2$. This constitutes a step forward the preceding results concerning these models~\cite{2rot_1,PhysRevLett.61.659}, where the exponential growth of the asymptotic energy was not found.

\begin{figure}
  \begin{center}
    \begin{tabular}{c}
     \hspace{0cm}\resizebox{85mm}{!}{\includegraphics{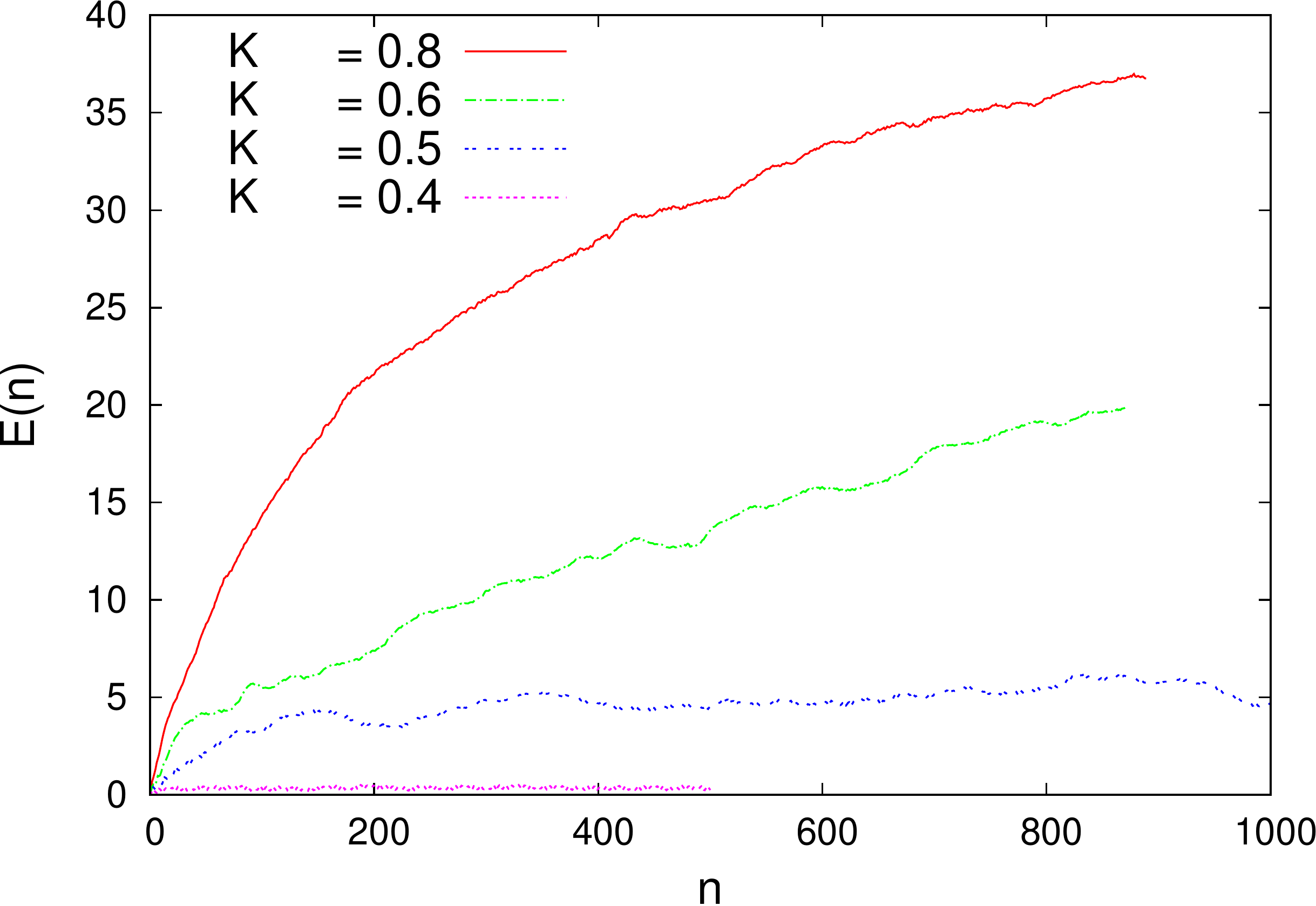}}
     \pgfputat{\pgfxy(-6.83,5)}{\pgfbox[left,top]{\small $/\kbar$}}\\
     \pgfputat{\pgfxy(-2.45,5.05)}{\pgfbox[left,top]{\small $/\kbar$}}
     \pgfputat{\pgfxy(-2.53,5.71)}{\pgfbox[left,top]{\small $/\kbar$}}
     \pgfputat{\pgfxy(-2.64,6.02)}{\pgfbox[left,top]{\small $/\kbar$}}
%
    \end{tabular}
  \end{center}
  \caption{Energy dynamics for $N=3$ and different values of $K$ obtained with TEBD algorithm. For the considered values of $K/\kbar$ we see a dynamical localization transition at $K/\kbar=0.6$. In all the cases we take $M\leq 8$, big enough so that the time traces have converged in $d$. $\kbar=400$, $\epsilon =-2$. }
  \label{3:fig}
\end{figure}

In FIG.~\ref{3:fig} we show results for $N=3$ obtained with the TEBD algorithm~\cite{Note_Davide}: 
we see that for $K/\kbar<0.6$ the energy tends to an asymptote and the system is dynamically localized; on the other side, for $K/\kbar>0.6$  the energy increases up to the bound imposed by the truncation dimension and the system is thus delocalized. While these results suggest the presence of a localization/delocalization transition, conclusive evidence may come only from an analysis of the localization properties of the Floquet states and the Floquet level spacing distribution which are the focus of the next subsections.

\subsubsection{Inverse participation ratio of the Floquet states}
%

Let us start by using the inverse participation ratio~\cite{Edwards_JPC72} (IPR) in the momentum basis: for a single Floquet state $\ket{\phi_\alpha}$ this object is defined as
\begin{equation}
  \mathcal{I}_\alpha\equiv\sum_{\bf m}|\left\langle{\bf m}\right.\ket{\phi_\alpha}|^4\,
\end{equation}
We will consider its average over the Floquet states in the truncated Hilbert space
\begin{equation}
  \overline{\mathcal I}_{M}=\frac{1}{M^N}\sum_\alpha\mathcal{I}_\alpha\,.
\end{equation}
If the Floquet states are localized in the momentum basis, this object does not scale with the local truncation dimension and tends to a limit $\overline{\mathcal I}_\infty$ which is finite for $M\to\infty$: each Floquet state has nonvanishing overlap only with a finite number of momentum eigenstates. In turn, if the Floquet states are delocalized in the momentum basis, we expect that the averaged IPR scales to 0 when the local dimension $d$ tends to $\infty$. 

\begin{figure*}
  \begin{center}
    \begin{tabular}{ccc}
%
     \hspace{-1cm}\resizebox{65mm}{!}{\includegraphics{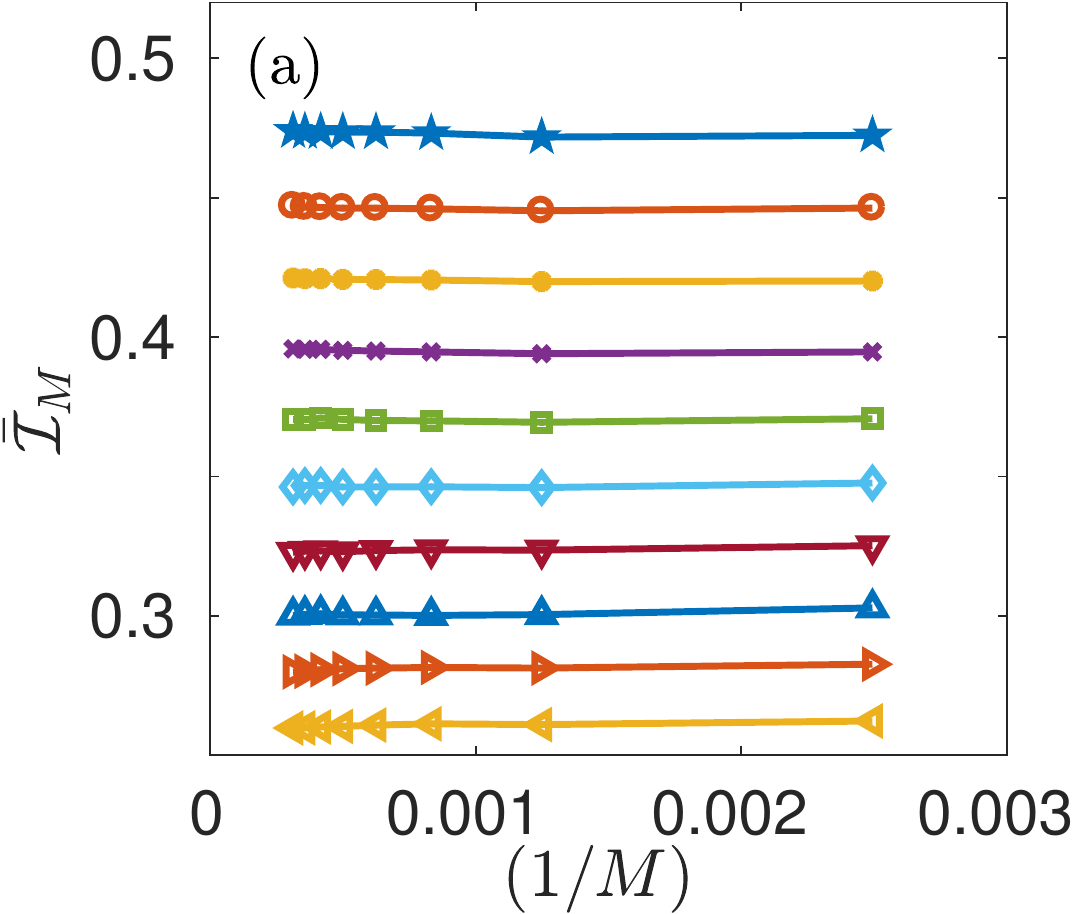}}&
     \hspace{0cm}\resizebox{65mm}{!}{\includegraphics{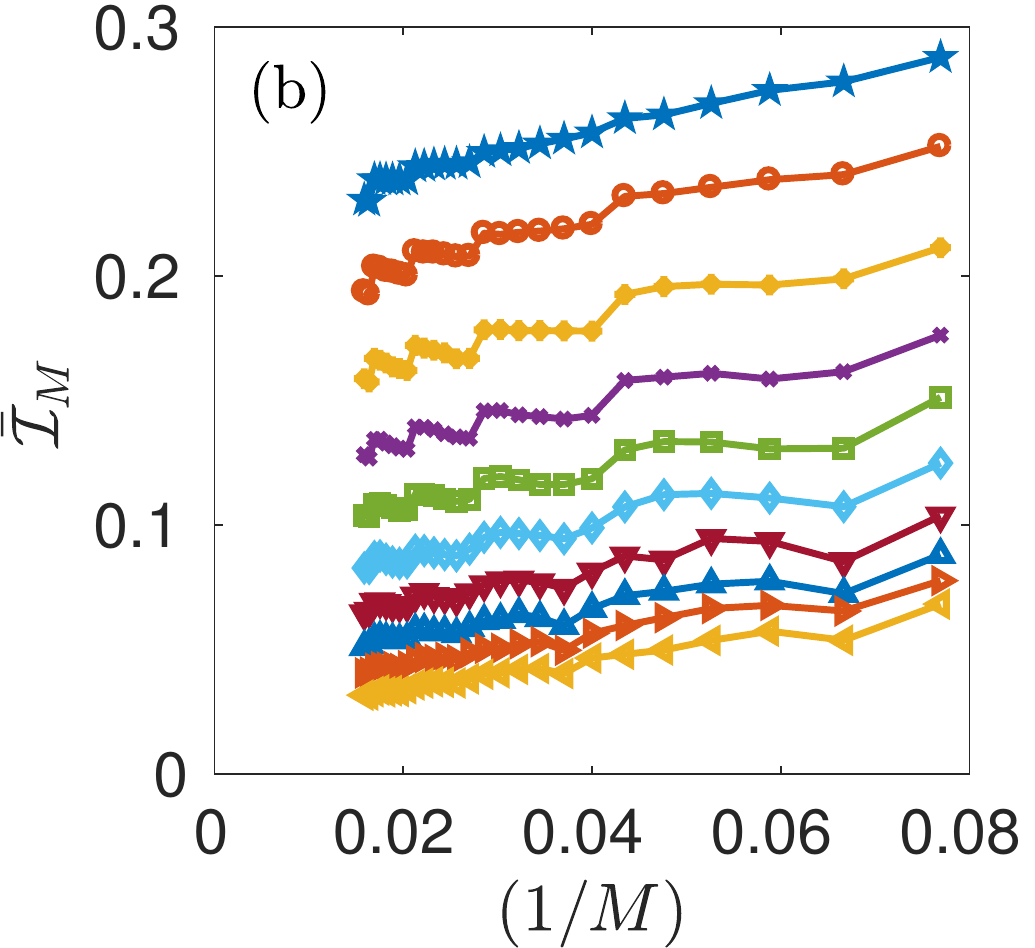}}&
     \hspace{0cm}\resizebox{65mm}{!}{\includegraphics{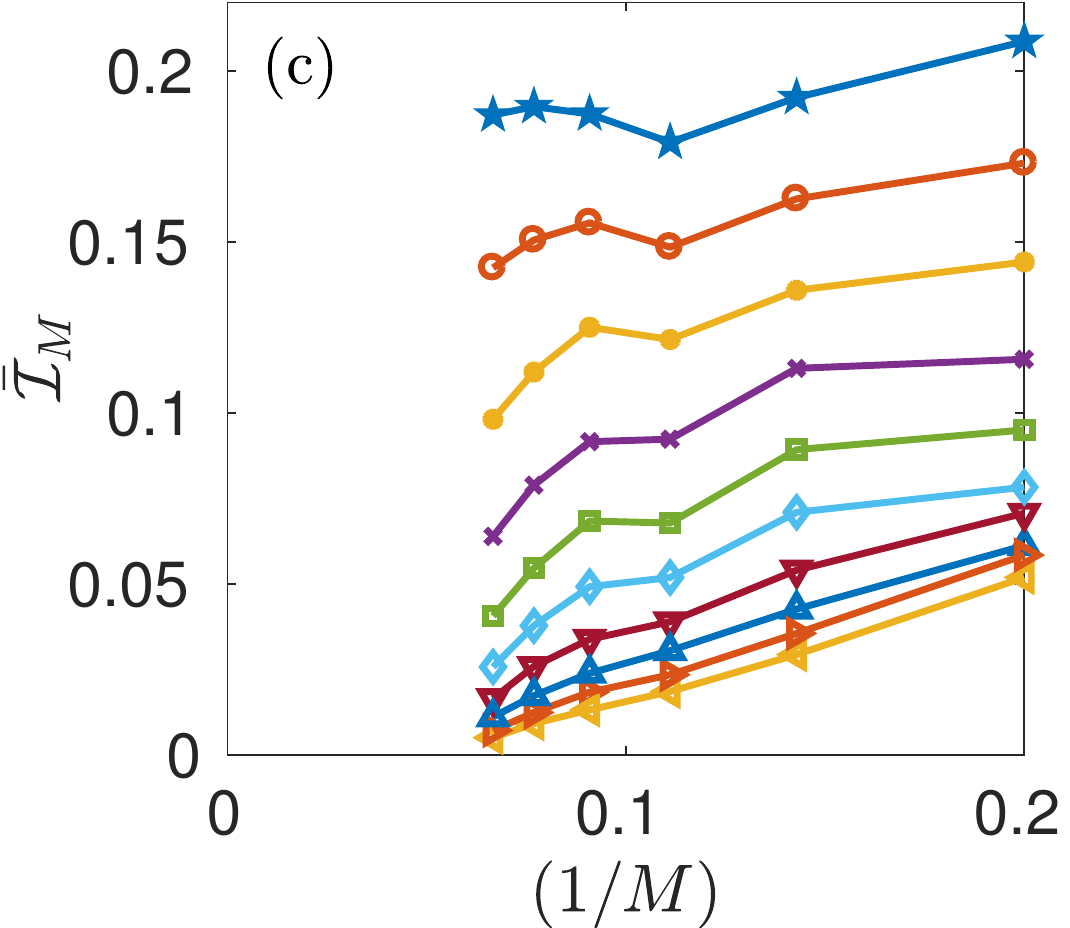}}\\
    \end{tabular}
    \hspace{0cm}\pgfputat{\pgfxy(-4.5,7.2)}{\pgfbox[left,top]{\resizebox{20mm}{!}{\includegraphics{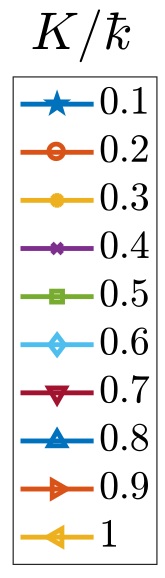}}}}
                 \resizebox{100mm}{!}{\includegraphics{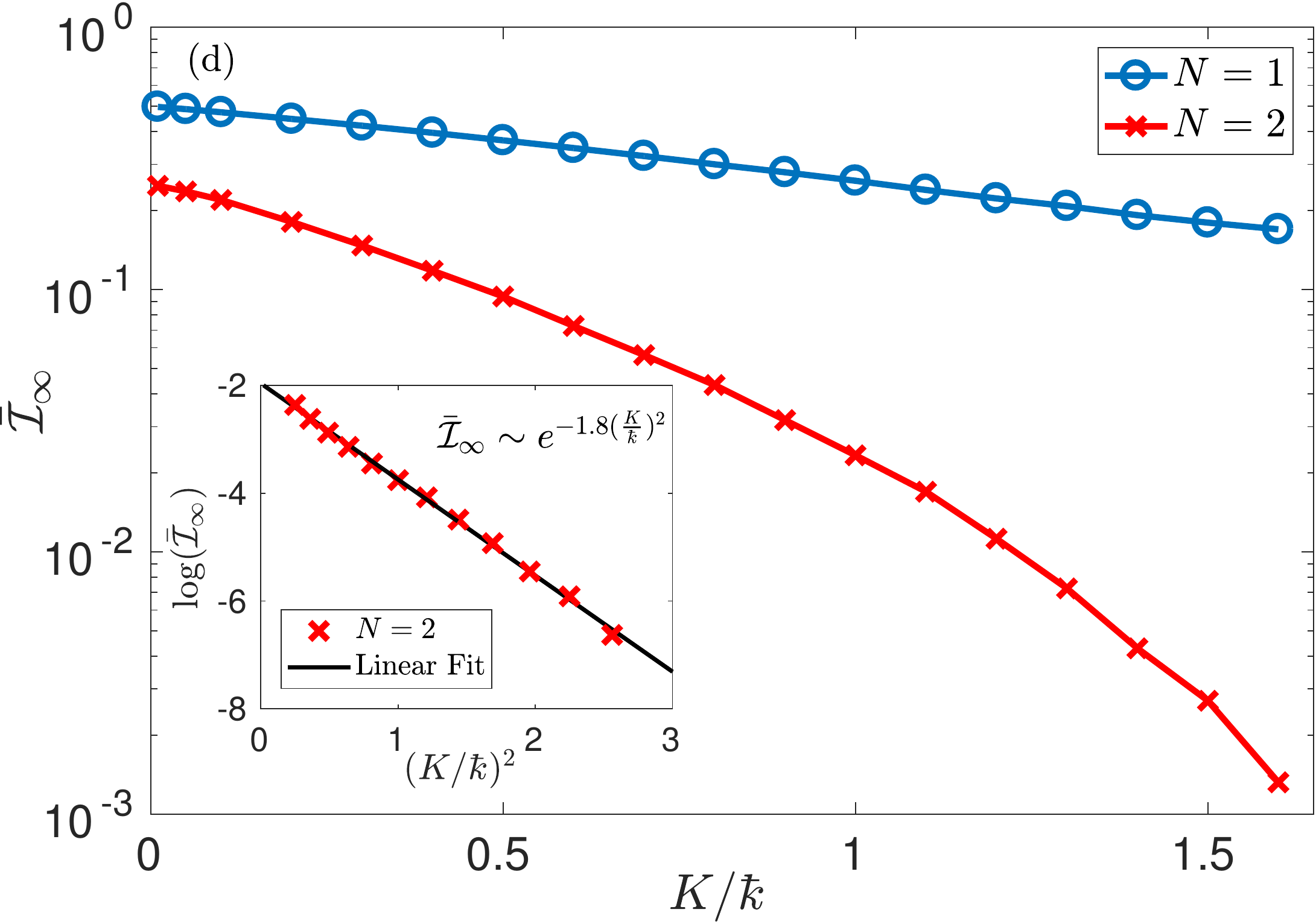}}
%
  \end{center}
  \caption{
  Plot of $\overline{\mathcal I}_M$ versus $1/M$ for $K/\kbar=0.1, 0.2, \dots, 1.0$ (see the legend in the lower left panel) in the three cases: $N=1$ (a), $N=2$ (b) and $N=3$ (c). Numerical parameters: $\epsilon=-2$ and $\kbar=400$.
 (d) Behavior of $\mean{I}_\infty$ vs $K/\kbar$ for $N=1$ and $N=2$; notice that its non-vanishing value is consistent with localization and that $\langle I\rangle_{\infty}$ decays exponentially fast in $K$.}
  \label{IPR2:fig}
\end{figure*}

We show numerical results in FIG.~\ref{IPR2:fig}: 
in panels (a), (b), (c) we plot $\overline{\mathcal I}_M$ versus $1/M$ in the cases $N=1$, $N=2$ and $N=3$ respectively. 
We report curves obtained for increasing values of $K=0.1, 0.2,\dots,1.0$: 
the values of $K/\kbar$ range from 0.1 to 1 spaced by intervals of 0.1 and the curves are always in a monotonously decreasing order in $K$ (see the color legend
in the lower left panel). 
In the case $N=1$ (a) the averaged IPR is almost constant in $1/M$ and tends to a finite $\overline{\mathcal I}_\infty$ for  $1/M\to 0$: 
the Floquet states are localized in the momentum basis. 
For $N=2$ (b) the (approximately linear) dependence on $1/M$ is more marked, but also in this case extrapolating to $1/d\to 0$ the limit is finite. 
In panel (d) we show the dependence of the limit $\overline{\mathcal I}_\infty$ on $K$, for $N=1$ and $N=2$. 
They are both obtained through linear interpolation of the data in the left and central upper panels. 
We see that they are both different from 0 but appear to decrease towards zero as $K$ increases: for $N=1$ the dependence is $\sim\nep^{-\alpha K}$ and for $N=2$ it is $\sim\nep^{-\beta K^2}$ (see the inset). 
For $N=3$, see panel (c), we cannot clearly see the localization/delocalization transition point, due to the limits on the values of $M$ which we can numerically consider. Nevertheless, for large $K$ we see delocalization: $\overline{\mathcal I}_M$ smoothly depends on $1/M$ and is consistent with a vanishing limit for $1/M\to 0$. In order to further explore the transition let us now turn to level spacings.
\subsubsection{Level spacing statistics}
Another tool we use to investigate the localized/delocalized behaviors of the system is the level spacing statistics.
The distribution of the Floquet level spacings $\mu_{\alpha+1}-\mu_\alpha$ (the
$\mu_\alpha$ are in increasing order) normalized by the average density of states gives information on
the integrability/ergodicity properties of the system~\cite{Haake:book,Berry_PRS76,Haake_ZPB86,Berry_LH84,Bohigas_PRL84,Poilblanc_EPL93}: 
if the distribution is Poisson the system is integrable, if it is Wigner-Dyson the system is ergodic. The level spacing distribution is therefore a probe for the system dynamics being integrable-like (regular) or ergodic.

This object is important to consider because there is a strict connection between ergodicity/regularity on one side and energy absorption/energy localization on the other, both in the classical and the quantum perspective.
Classically a system is ergodic if all the trajectories uniformly explore the accessible part of the phase space. If energy is conserved, this part is the energy shell: as a consequence the system thermalizes (time averages equal microcanonical averages). If energy is not conserved (as in a periodically driven system), ergodicity implies uniform exploration of all the phase space and then thermalization at $T=\infty$. Therefore, if the energy spectrum is unbounded, ergodicity is strictly connected with infinite energy absorption~\cite{NOte7}. In ergodic quantum systems the same phenomena result from the eigenstates of the dynamics being locally equivalent to the microcanonical ensemble: this is a consequence of them behaving as the eigenstates of a random matrix (eigenstate thermalization -- see for instance~\cite{Deutsch_PRA91,Sred_PRE94,Rigol_Nat,Bohigas_PRL84}). In the kicked case the Floquet states are locally equivalent to the completely mixed density matrix and this fact gives rise to $T=\infty$ thermalization~\cite{Rigol_PRX14,Lazarides_PRE14,Ponte_AP15,Rosch_PRA15,Russomanno_EPL15}. As a consequence, they are extended in any basis of ``simple'' states: the IPR evaluated in that basis will vanish with the dimension of the Hilbert space, as observed in the subsection above for the case of the momentum basis.

On the opposite, in the case of classical dynamical localization, there are constraints for the dynamics which forbid the system to uniformly explore the phase space and thermalize. This is the case of integrable systems which have an extensive amount of integrals of motion with vanishing Poisson brackets~\cite{Berry_regirr78:proceeding,Arnold:book}. 
For instance, in the case of a classical kicked rotor with small amplitude kicking, a significant portion of the phase space behaves regularly, giving rise to dynamical localization. From the quantum point of view, the trajectories being constrained in a small portion of the phase space reflect in the eigenstates of the dynamics not being random superpositions of elements of some local basis, but being localized in this basis.
Therefore, we expect to see signatures of integrable behavior also in the case of quantum dynamical localization, especially in the properties of the level spacing distribution which should be Poisson like. 

\begin{figure}
  \begin{center}
    \begin{tabular}{c}
     \hspace{-0.2cm}\resizebox{\columnwidth}{!}{\includegraphics{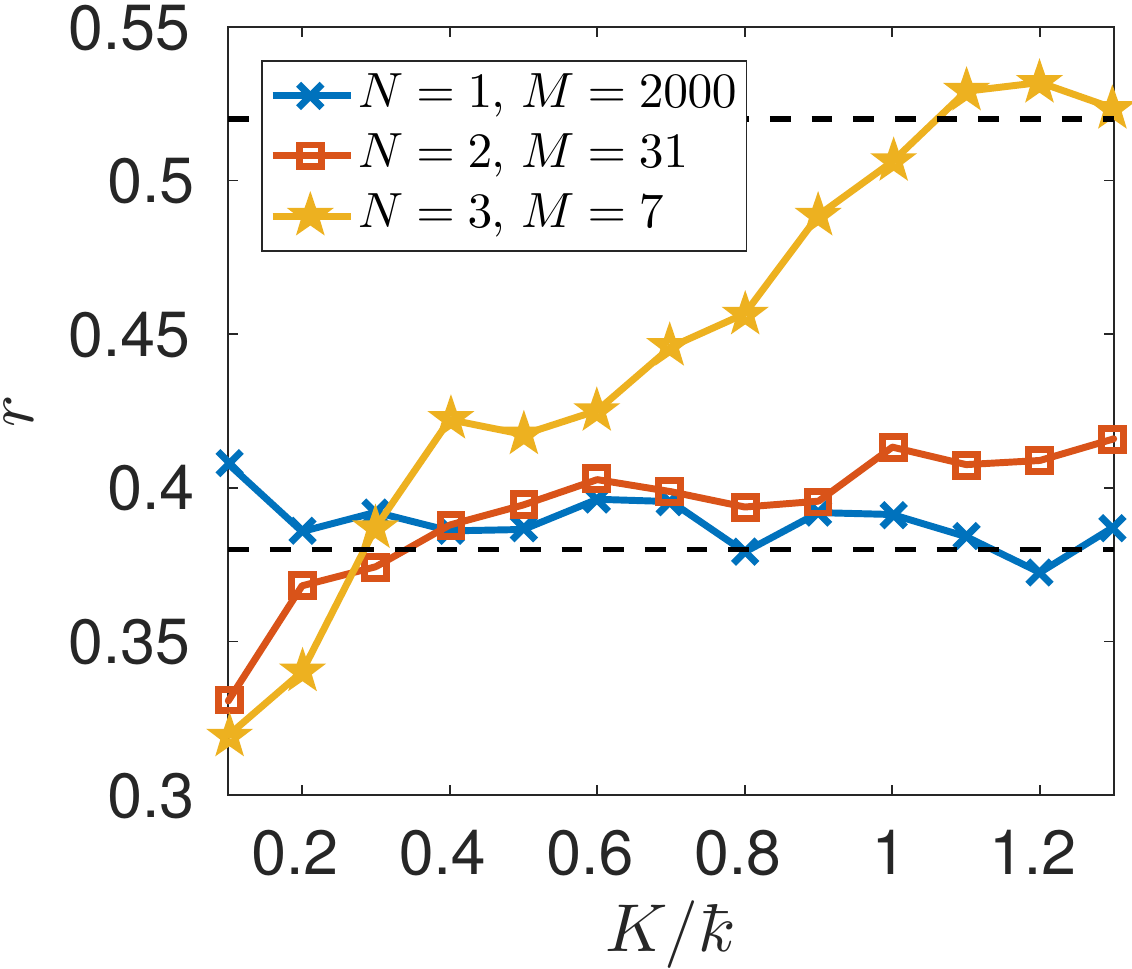}}\\
    \end{tabular}
  \end{center}
  \caption{Level spacing ratio averaged over the whole Floquet spectrum vs. $K/\kbar$: dynamical localization corresponds to Poisson-like behavior. The 
  lower dashed line is the Poisson value $r=0.386$ - corresponding to integrability - while the upper one the Wigner-Dyson one $r=0.5295$, corresponding to ergodicity.
  Numerical parameters $\kbar=400$, $\epsilon =-2$.}
  \label{averello:fig}
\end{figure}

In order to probe the integrability/ergodicity properties through the level spacing distribution, we consider the so-called level spacing ratio
$r_\alpha$. If we define $\delta_\alpha=\mu_{\alpha+1}-\mu_\alpha$, we have
\begin{equation}
  0\leq r_\alpha \equiv \frac{\min\left\{\delta_\alpha,\delta_{\alpha+1}\right\}}{\max\left\{\delta_\alpha,\delta_{\alpha+1}\right\}}\leq 1\,.
\end{equation}
The different level spacing distributions are characterized by a different value of the average $r\equiv\mean{r_\alpha}$ over the distribution. From
the results of Ref.~\cite{oganesyan2007localization}, we expect $r=0.386$ if the system behaves integrably and the distribution is Poisson; on the other side, if the distribution is Wigner-Dyson and the system behaves ergodically, then $r=0.5295$. Being the Hamiltonian Eq.~\eqref{H_a:eqn} symmetric under on-site inversion ($\hat{p}_j\to-\hat{p}_j$, $\hat{\theta}_j\to-\hat{\theta}_j$) and under global reflection ($\hat{p}_j\to\hat{p}_{L-j+1}$, $\hat{\theta}_j\to\hat{\theta}_{L-j+1}$) we need to evaluate the level spacing distribution and the corresponding  $r$ only over Floquet states in one of the symmetry sectors of the Hamiltonian~\cite{Haake_ZPB86}.
We show numerical results obtained through exact diagonalization in FIG.~\ref{averello:fig}. 
We see that, for $N=1$, $r$ is always near to the Poisson value: this is consistent with the system being always dynamically localized. 
For $N=2$, $r$ is close to the Poisson value in the interval where we are able to see dynamical localization in FIG.~\ref{2:fig}: 
also in this case our hypothesis of connection between the integrable behavior of the system and the energy localization is confirmed. 
Around $K/\kbar=1.5$, $r$ deviates from the Poisson value: the momentum localization length increases exponentially with $K$ and at a certain point it is larger than the truncation dimension $M$. 
When $N=3$, although we can only numerically consider a quite small value of $M$, we see that $r$ increases with $K$ and eventually sets to the Wigner-Dyson value.
There is indeed a crossover between Poisson and Wigner-Dyson; we see that $K^*/\kbar\simeq 0.6$, the localization/delocalization transition point seen through the energy dynamics in FIG.~\ref{3:fig}, falls in the intermediate region, at a value where $r$ is near to Poisson. In the limit $M\to\infty$, most probably $r$ tends to the Poisson value for $K<K^*$, but we do not know if the crossover develops into a clear-cut transition. If some intermediate region persisted in this limit, localized and delocalized Floquet states would appear in different parts of the spectrum (though not coexisting at the same quasienergy).
Something similar happens in classical chaotic systems, where regular and chaotic trajectories exist together when the system is in the transition region between integrability to ergodicity. Nevertheless, when $N>1$, the system eventually thermalizes also in the transition region~\cite{konishi} (this is a manifestation of the Nekhoroshev theorem and the Arnold diffusion~\cite{Arnold_Avez:book}). Of course further research is needed to clarify this point.
%
%
%
\section{Absence of localization for $N\to\infty$} \label{thermodyno:eqn}
In this section we discuss the behavior of the coupled rotors model in the thermodynamic limit $N\to\infty$ and  show that the mapping introduced for finite $N$ in the previous section is valid also in this limit.
Applying the scaling theory of localization~\cite{Licciardello} we find in Subsection~\ref{noloc_inf:sec} 
that the localization/delocalization transition of a disordered $N-$dimensional lattice disappears for $N\to\infty$: the system is always delocalized in this limit. 

In Subsection~\ref{mean_field1:sec} we study numerically the behavior of the kicked rotors for $N\to\infty$ using a time-dependent mean field approach. This approach is exact when the coordination number is infinite; this can happen, for instance, when the interactions in the model of Eq.~\eqref{H_a:eqn} are  infinite-range and we are in the thermodynamic limit. 
By changing the kick amplitude and the interaction coupling we find two regions in the parameter space, one in which the dynamics is diffusive and one in which it is subdiffusive. 
Focusing on the time evolution of the mean field parameter, which in the MF approximation is controlling the effective kicking strength, we study the relation between its behavior and the subdiffusive/diffusive growth of the kinetic energy. 
Considering the average over an ensemble of different initial conditions, we see that the mean field parameter behaves as a non stationary signal, with a variance decreasing as a power law. 
We also consider the spectral properties of the mean field parameter: the parameter itself and its time correlator exhibit  power law behaviors at small frequencies when the dynamics is subdiffusive. The exponents of the power laws decrease as the kicking strength is increased and completely disappear when the dynamics is diffusive: in this case the power spectra are completely flat.

 

\subsection{Delocalization in the infinite-dimensional Anderson model} \label{noloc_inf:sec}
We start our discussion showing that there is no Anderson localization in a $N$-dimensional disordered lattice like the one in Eq.~\eqref{eigenstat:eqn} when the dimension $N$ tends to infinity. To that purpose we use the scaling theory of localization introduced in Ref.~\cite{Licciardello} which we briefly summarize to fix the notation. Consider a system with conductivity $\sigma$ and focus on the properties of the dimensionless conductance $g=\frac{\hbar}{e^2}L^{N-2}\sigma$. We make the assumption that $g$ only depends on the scale $L$ at which the system is probed and on the dimensionality $N$ and we start at some scale $L_0$ where the dimensionless conductivity is $g_0$: we see how $g$ flows as $L$ is increased. For that purpose it is crucial to focus on the properties of the logarithmic derivative $\beta(g)\equiv\frac{\ud\log g}{\ud \log L}$ and in particular on its dependence on $g$: knowing the form of $\beta(g)$ and integrating this flow equation, the bulk behavior for $L\to\infty$ is obtained. It is possible to find the behavior of $\beta(g)$ in the limits $g\ll 1$ and $g\gg1$. When $g\ll 1$ there is Anderson localization, and the conductance obeys the law $g(L)\sim A\nep^{-L/\xi}$ for some localization length $\xi$: this relation implies
\begin{equation}  \label{anderson_loc:eqn}
  \beta(g)=\log g + {\rm const}\,.
\end{equation}
 In this limit, $\beta(g)$ versus $\log g$ is a line whose slope is independent of the localization length and equals 1. In the opposite limit of $g\gg 1$ there is Ohmic conductivity, $\sigma$ does not depend on $L$ and $\beta(g)=N-2$. The question is how to interpolate between these two limits. One can show~\cite{Licciardello,Rammer:book,Lee_RMP85} that, because of the quantum corrections to the Ohm's law, in the limit of large $g$ it is
\begin{equation} \label{weak_scaling:eqn}
  \beta(g)\simeq N-2-\frac{C(N)}{g}\,,
\end{equation}
for some $C(N)$ depending on the dimension. 
Connecting this large-$g$ behavior with the small-$g$ linear one Eq.~\eqref{anderson_loc:eqn} in a continuous and derivable way, one gets a $\beta(g)$ which is always monotonously increasing. 
(The physical assumption behind this connection is that at some point the ``weak localization'' due to the quantum corrections to the Ohm's law becomes the strong Anderson localization). 
Therefore, we always have  $\frac{\ud\beta (g)}{\ud \log g}>0$. This gives rise to interesting consequences. 
For $N\leq 2$ we find as a consequence that  $\beta(g)=\frac{\ud\log g}{\ud \log L}<0$ for all $g$: 
when larger and larger values of $L$ are considered, whichever are the initial values $L_0$ and $g_0$, they always flow towards small values of $g$, the ones corresponding to Anderson localization. 
If instead $N>2$, there is some value $g_N^*$ where $\beta(g_N^*)=0$. For $g>g_N^*$ we have $\beta(g)>0$ and $g<g_N^*$ implies $\beta(g)<0$.
Therefore, if $g_0<g_N^*$ the system flows towards small values of $g$ for $L\to\infty$ and there is Anderson localization in the bulk; if instead $g_0>g_N^*$ the flow moves towards large values of $g$ and there is an Ohmic behavior. Therefore for $N>2$, the bulk of the system undergoes a localization/delocalization transition. 
We have observed exactly this phenomenon in Section~\ref{mapping:sec} for the model with three rotors mapped over the $N=3$-dimensional disordered lattice Eq.~\eqref{eigenstat:eqn}: 
in this case $K$ played the role of $g_0$. Now we would like to explore the behavior of $g_N^*$ in the limit $N\to\infty$. 
To that purpose, we study the behavior of the conductivity: its quantum corrections to the Ohmic behavior are~\cite{Rammer:book,Lee_RMP85}
\begin{eqnarray}
  \delta\sigma(L)&=&-\frac{e^2}{\pi\hbar}\int_{1/L}^{1/l}\frac{\ud^NQ}{(2\pi)^N}\frac{1}{Q^2}\nonumber\\
  &=&-\frac{2e^2}{\pi\hbar(2\pi)^N}\frac{S_{N-1}}{N-2}\left(\frac{1}{l^{N-2}}-\frac{1}{L^{N-2}}\right)\,,
\end{eqnarray}
where 
\begin{equation} \label{sphera:eqn}
  S_{N-1}=\frac{2(\pi/2)^{N/2}}{\Gamma(N/2)}
\end{equation}
 is the measure of the $N-1$-dimensional unit sphere and $l$ is the classical mean-free-path in the disordered potential (its precise value is not important because it will disappear in the next formulae). Using that $g(L)=\frac{\hbar}{e^2}L^{N-2}(\sigma(\infty)+\delta\sigma(L))$, we easily find that $\beta(g)$ has the form given in Eq.~\eqref{weak_scaling:eqn} with
\begin{equation}
  C(N)=\frac{(\pi/2)^{N/2}}{(2\pi)^N\Gamma(N/2)}\,.
\end{equation}
Connecting Eq.~\eqref{weak_scaling:eqn} in a continuous and derivable way with the Anderson-localized behavior Eq.~\eqref{anderson_loc:eqn} valid at small $g$, we find that the critical value $g_N^*$ is given by
\begin{equation}
  \log g_N^*=-N+3+\log\left(\frac{2}{\pi}\frac{1}{(2\sqrt{2\pi})^N}\frac{1}{\Gamma(N/2)}\right)\,,
\end{equation}
for $N$ large enough. For $N\gg 1$, using the Stirling approximation for the Gamma function, we find
\begin{eqnarray}
  \log g_N^*&=&-N+3+\log(2/\pi) - N\log(2\sqrt{2\pi}) \nonumber\\
  &-& \left(\frac{N}{2}+1\right)\left[\log\left(\frac{N}{2}+1\right)-1\right]\,.
\end{eqnarray}
We see therefore that $\lim_{N\to\infty}g_N^*=-\infty$: for $N\to\infty$ the critical value of $g$ is zero and therefore the system is always delocalized.


\subsection{Mean field approach}\label{mean_field1:sec}
To study directly the large $N$ limit we apply the mean field approximation which is exact for infinite coordination number or infinite range interactions. We will focus on the latter case. 
Henceforth throughout this subsection we will consider the Hamiltonian Eq.~\eqref{H_sr:eqn} with  $\epsilon_{ij}=\frac{\epsilon}{N-1}$. We then perform a mean field {\em Ansatz}: starting from a factorized state, we assume that the system remains factorized during the whole time evolution. Corrections to this behavior turn out to be negligible in the limit $N\to\infty$. The many body initial  state we are considering is therefore of the form 
\begin{equation} \label{approx_stato:eqn}
  \ket{\Psi_{\rm MF}(0)}=\prod_i\ket{\psi_i(0)}\,.
\end{equation}
%
We assume translation invariance, therefore all the initial $\ket{\psi_i(0)}$ are equal to some $\ket{\psi(0)}$, and all of them evolve to the same single-site state. We define this single-site state just before the $n$-th kick as $\ket{\psi(n)}$: the corresponding many-body state is the tensor product of $N$ copies of this state. In this way, we can describe the dynamics of the system via an effective single particle Hamiltonian containing a time modulation of the  kick:
\begin{align}\label{mf_H:eqn}
\hat{H}_{MF} = &\ \frac{1}{2} \hat p^2+K\sum_{n=-\infty}^{+\infty}\delta (t-n) \bigg[  \cos \hat\theta -\\\nonumber 
&\frac{\epsilon}{2} \left( \psi_{MF}(n)\, \mathrm{e}^{-i \hat\theta} +\mathrm{h. c.}\right)\bigg]\,,
\end{align}
%
where we have defined the complex mean-field parameter
\begin{align} \label{parametrazzi:eqn}
\psi_{MF}(n) &\equiv \langle \psi(n)\,| \mathrm{e}^{i \hat\theta}\,   |\, \psi(n)\rangle\,,
\end{align}
%

This description is exact for infinite range interactions in the thermodynamic limit. To see this rewrite the interaction term in Eq.~\eqref{H_sr:eqn} as
\begin{align} \label{eqn:formulona}
&V(\hat{\boldsymbol\theta})=- \frac{\epsilon}{2(N-1)}\sum_{i \neq j} \bigg[\mean{\mathrm{e}^{i \hat\theta_i}}_n\, \mathrm{e}^{-i \hat\theta_j} +\mathrm{h. c.}\bigg] \\ \nonumber
&+ \frac{\epsilon}{2}\sum_{i}\left|\mean{\mathrm{e}^{i \hat\theta_i}}_n\right|^2
-\frac{\epsilon}{2(N-1)}\sum_{i\neq j}\left[\hat\delta_j\hat\delta_i+\hat\chi_j\hat\chi_i \right]\,,
\end{align}
%
where $\hat\delta_i\equiv\cos\hat\theta_{i}-\mean{\cos \hat\theta_i}_n$ and $\hat\chi_i\equiv\sin\hat\theta_{i}-\mean{\sin \hat\theta_i}_n$ and $\mean{}_n$ is the expectation value over the exact solution of the Schr\"odinger equation. 
Imposing translation invariance,  the first sum gives the single particle mean-field potential of Eq.~\eqref{mf_H:eqn}. The second sum is in turn a time dependent c-number term that can be neglected. The third sum contains terms in the form $\hat\delta_j\hat\delta_i$ and $\hat\chi_j\hat\chi_i$ with $i\neq j$: their expectation value $\langle	\hat\delta_j\hat\delta_i\rangle$ at time $n$ is a spatial connected correlator for the cosine ($\langle	\hat\chi_j\hat\chi_i\rangle$ is the same for the sine). These connected correlators vanish in the thermodynamic limit for each $i$ and $j$ (see Appendix~\ref{mf_calc}): more precisely, 
we explicitly compute $\mean{\cos\hat\theta_i\,\cos\hat\theta_j}_n$ and show that it can be factorized up to corrections which
vanish at the leading order as $O(n/N)$, if the state at time $n=0$ is separable. 
Therefore the expectation value of the sum of  $\hat\delta_j\hat\delta_i$ grows in a non extensive way ($\sim\sqrt{N}$) and therefore is negligible in the limit $N\rightarrow\infty$.
We see therefore that spatial correlations vanish for $N\rightarrow\infty$: the {\em Ansatz} that we made above is valid and therefore the separability of the initial state 
is preserved during the evolution. This definitively allows us to study our system via the effective mean-field single particle model described by the Hamiltonian in Eq.~(\ref{mf_H:eqn}).

It is convenient at this point to express the initial wave function in the momentum basis: in the angle representation we have
%
\begin{equation}\label{wf_comp:eqn}
  \bra{\theta}\psi(0)\rangle=\sum_{m=-\infty}^{+\infty}a_m\mathrm{e}^{im\theta}\,,
\end{equation}
and average over many random initial conditions with a fixed kinetic energy (the average symbol is $\overline{(\cdot)}$). The initial conditions are
obtained by applying one kick to  the zero-momentum state and then randomizing the phases of the amplitudes in the momentum basis.
%
We consider initial states such that $a_m=a_{-m}$: it follows that 
$\psi_{MF}(n)$ is real and that
the evolution operator over one period at time $n$ can be written as \cite{NOte6}
\begin{align} \label{quant_Uf2}
&\hat U_{MF}(n)= \mathrm{e}^{-\frac{i}{\kbar}\frac{\hat p^2}{2} } \mathrm{e}^{-i\frac{K}{\kbar}[ 1-\epsilon\ \psi_{MF}(n)]\cos \hat\theta}\,.
\end{align}
 $\hat U_{MF}(n)$ depends on the state at time $n$ through the mean field parameter $\psi_{MF}(n)$, which is evaluated according to the prescription given in Eq.~\eqref{parametrazzi:eqn}. By iterating this procedure we generate the dynamics of the system starting from the initial state Eq.~\eqref{wf_comp:eqn}. 

As a result of the mean field approach, the many-rotors model is effectively described by a single rotor with a time dependent kicking strength given by
\begin{equation} \label{def_K_t}
K(n)=K[1-\epsilon\psi_{MF}(n)]\, .
\end{equation}
Below we focus on the analysis of the dynamics of the following quantities:
\begin{enumerate}
  \item[\textit{1.}] the kinetic energy $E(n) = \overbar{\langle\hat{p}^2\rangle_n}/2$ averaged over the initial conditions (for each evolution we define $\langle\hat{p}^2\rangle_n\equiv\bra{\psi(n)}\hat{p}^2\ket{\psi(n)}$);
  \item[\textit{2.}] the power spectrum $P(\omega)=|\widetilde{\psi}_{\omega}|^2$, where $\widetilde{\psi}_{\omega}$ are the Fourier coefficients of $\overbar{\psi_{MF}(n)}$;
  \item[\textit{3.}] the power spectrum $P_{ac}(\omega; n_0)=|\widetilde{c}_{MF}(\omega;n_0)|^2$, where we define the  correlator
\begin{align}\label{acf_def}
& c_{MF}(k;n_0) =\\ \nonumber
& \overbar{\psi_{MF}(n_0)\psi_{MF}(n_0+k)} - \overbar{\psi_{MF}(n_0)}\,\overbar{\psi_{MF}(n_0+k)};
\end{align}
  \item[\textit{4.}] the variance of the mean field parameter, defined as 
\begin{equation}
\sigma_{MF}(n)=\overbar{\psi_{MF}(n)^2}- \left(\overbar{\psi_{MF}(n)}\right)^2.
\end{equation}
\end{enumerate}
The first quantities characterizes the energy dynamics of the system and its ergodicity properties, while the others analyze the mean field parameter. 
As discussed in the previous section, also in this case 
the local Hilbert space is infinite dimensional (see Eq.~\eqref{wf_comp:eqn}) and a truncation is therefore necessary.
The truncation dimension $M$ varies according to the parameters $K$ and $\epsilon$ and to the length of the simulation; 
it is chosen such that higher momentum states are not involved in the evolution. 
The evolution operator defined in Eq. (\ref{quant_Uf2}) is factorized in two parts: one is diagonal in the momentum basis and the other in angle representation. We generate the time evolution over one period  by applying separately the kinetic and the kick part to the wave-function. We work in the former case in the momentum basis, in the latter in the angle one.

\begin{figure*}
\centering
\includegraphics[width=12cm]{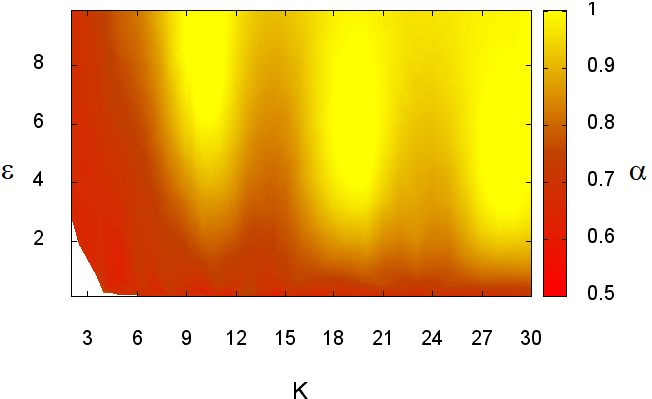}
\caption{\label{fig:pow_law_exp}The power law exponent for $E(n)$ growth is plotted against the kick strength $K$ and coupling amplitude $\epsilon$. The subdiffusive region (red, dark one) and the diffusive one (yellow, light one) can be distinguished (color online). The region in the left-bottom corner is not plotted since a stable growth regime does not start within the simulation time length. We put $\kbar = 2.89$ since this value was used in an experimental realization of a kicked rotor with ultracold atoms \cite{PhysRevA.80.043626}.}
\end{figure*}

\subsubsection{Kinetic energy $E(n)$}

From the simulations  we find that $E(n)$ grows in time according to a power law $n^\alpha$, with $\alpha$ depending on $K$ and $\epsilon$:
this dependence is shown in FIG.~\ref{fig:pow_law_exp} in which the exponent $\alpha$ is plotted in the $(K, \epsilon)$ plane. 

\begin{figure}
\hspace{0.0cm}\resizebox{\columnwidth}{!}{\includegraphics{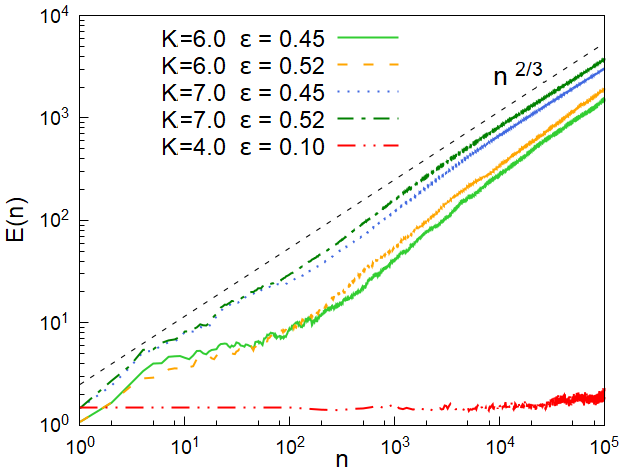}}
\caption{\label{fig:quant_subdiff}Time evolution of $E(n)$ is plotted together with the curve $n^{2/3}$ as a guide to the eye: the power law growth of $E(n)$ starts at different times but is characterized by an exponent $\alpha$ with a value in the interval $[0.6,\,0.7]$. For $K=4.0$ the energy starts growing at $t\sim 10^6$. Numerical parameters: $\kbar=2.89$.}
\end{figure}

We can distinguish two regions (see FIG.~\ref{fig:pow_law_exp}): the red, dark one (color online) in which $E(n)$ grows subdiffusively and the yellow, light one in which diffusion is observed. Subdiffusion is an effect purely due to the quantum nature of the system since the classical counterpart always exhibits normal diffusion in all the $(K, \epsilon)$ plane (see Appendix~\ref{classical:sec}). The exponent $\alpha$ is almost uniform in all the subdiffusive (red) region in the parameter space with values between $0.6$ and $0.7$. The transition from the subdiffusive behavior to the diffusive one is characterized by a variation of the power law exponent.
Some energy time-traces corresponding to different values of $K$ and $\epsilon$ are shown in FIG.~\ref{fig:quant_subdiff}.
The subdiffusive regime starts at a time $t$ which increases by lowering the values of $K$ and $\epsilon$: 
during the transient the energy first keeps constant,  then it starts growing until it reaches the $n^\alpha$ regime.
For certain value of $(K,\epsilon)$ (e.g. $K=4$ and $\epsilon=-0.1$) we do not see the start of either diffusion or subdiffusion within our simulation time ($\simeq 10^6$): the trend appears however to rule out localization but rather suggest that $t \geq 10^6$.

\subsubsection{Power spectrum $P(\omega)$}
In the study of $P(\omega)$ we distinguish its behavior at low and high frequencies: at low frequency we observe either a power law decay in $\omega$ 
or a constant power spectrum depending on whether $(K,\epsilon)$ are in the subdiffusive or diffusive region. 
\begin{figure*}
  \begin{center}
    \begin{tabular}{cc}
%
    \hspace{-0.5cm}\resizebox{9.0cm}{!}{\includegraphics{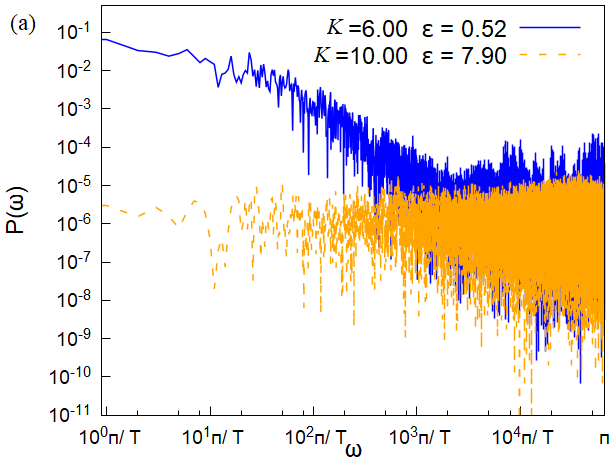}}&
    \hspace{0.5cm}\resizebox{9.0cm}{!}{\includegraphics{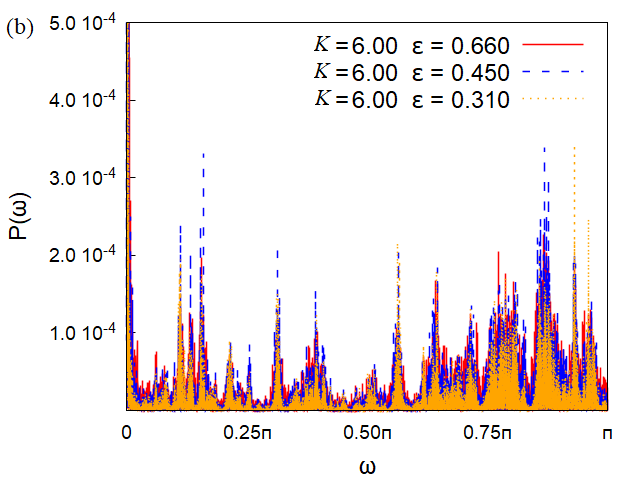}}
    \end{tabular}
  \end{center}
  \caption{\label{fig:pow_spectr_psi2}{\rm (a)} The low frequencies behavior of  $P(\omega)$  is plotted for a case in which the dynamics is diffusive (dashed line) and another in which it is subdiffusive manifests (continuous line). $\mathcal{T}=65536$ is the length of the time interval which has been used to compute the Fourier transform; it coincides with the number of frequencies which has been considered. {\rm (b)} High frequencies behavior of $P(\omega)$ for a fixed value of $K$ and different values of $\epsilon$: the positions of the peaks almost coincide. The norm of the power spectrum has been normalized to unity in order to enhance the visibility of the peaks within the same order of magnitude. In the simulations $\kbar= 2.89$.
}
  \end{figure*}

\begin{figure*}
  \begin{center}
    \begin{tabular}{cc}
    \hspace{-0.5cm}\resizebox{9.cm}{!}{\includegraphics{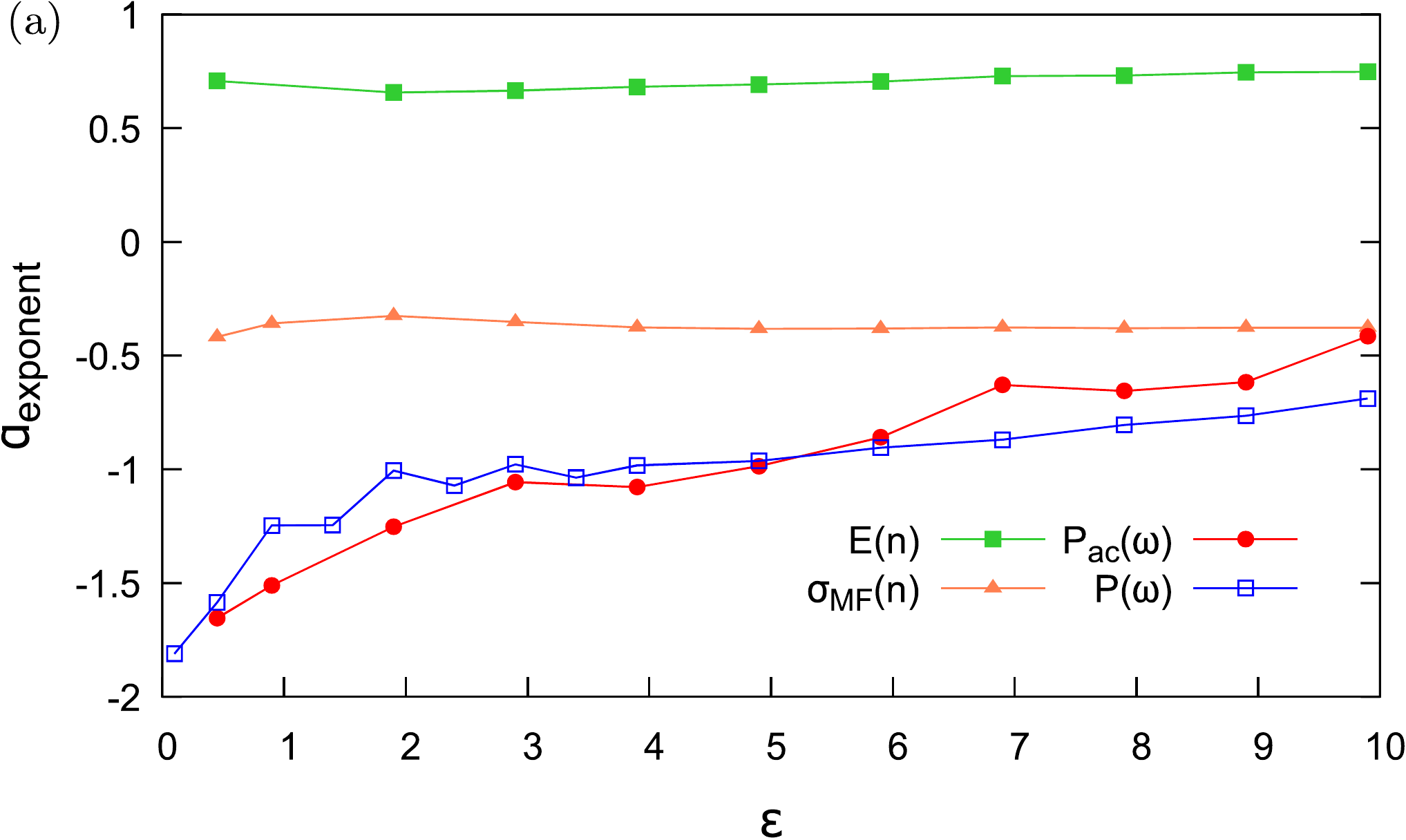}}&
    \hspace{0.5cm}\resizebox{9.cm}{!}{\includegraphics{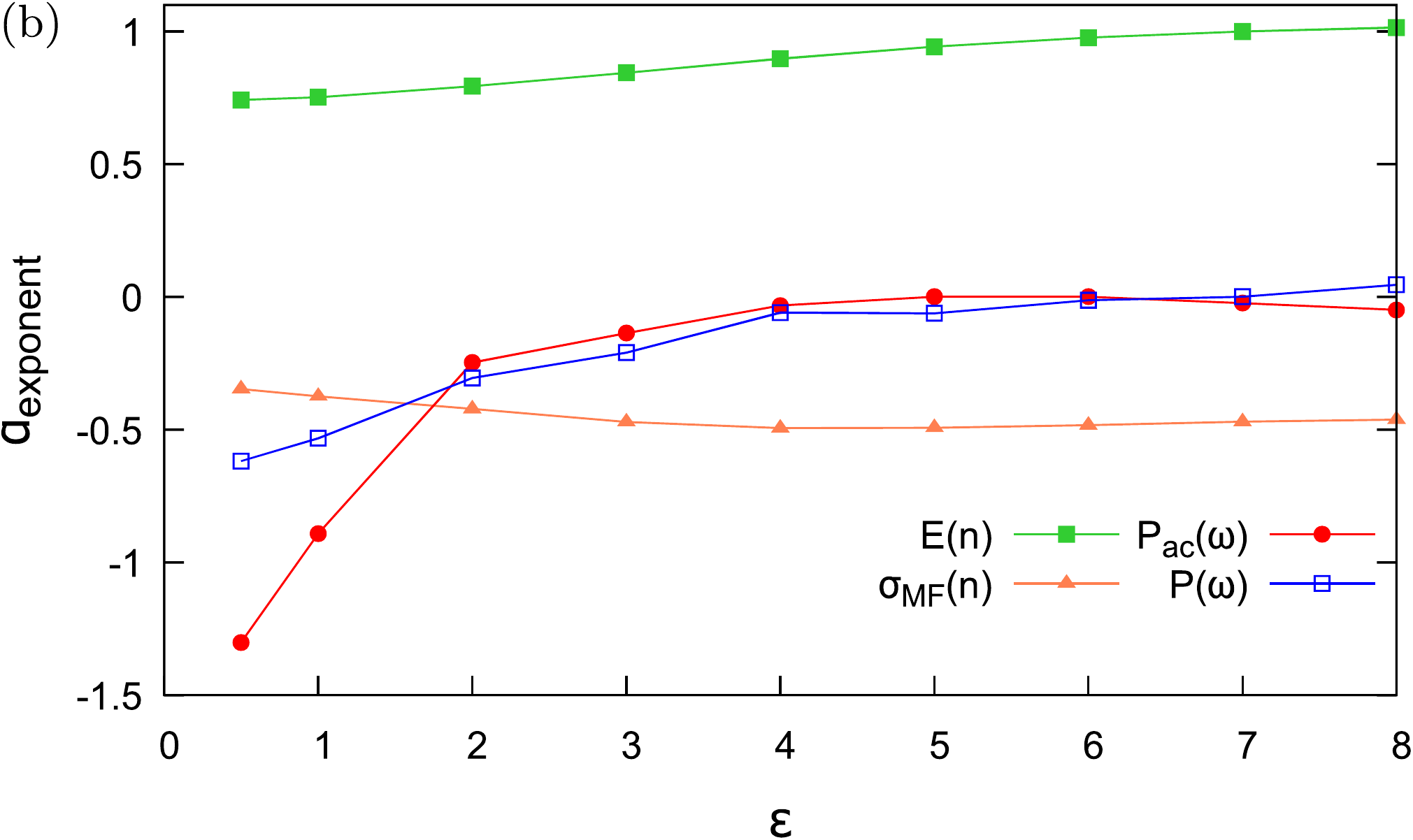}}
    \end{tabular}
  \end{center}
  \caption{The power-law exponents for $E(n)$ and $\sigma_{MF}(n)$ in the time domain and for $P(\omega)$ and $P_{ac}(\omega)$ in frequency domain are plotted. (a) $K=4.0$: the exponents relative to $E(n)$ and $\sigma_{MF}(n)$ (green square and orange triangles respectively) are uniform in $\epsilon$; the power laws exponents of the power spectra (red circles for $P(\omega)$ and empty blue ones for $P_{ac}(\omega)$) increase with $\epsilon$.  (b) $K=11.0$: the dynamics of the system passes from subdiffusive to diffusive when $\epsilon$ is increased. The exponents of the power laws of $P(\omega)$ and $P_{ac}(\omega)$ vanish when the dynamics is diffusive.  In the simulations $\kbar= 2.89$.
}
  \label{fig:plk411}
\end{figure*}
%
%
The low frequencies behavior is shown in panel (a) of FIG.~\ref{fig:pow_spectr_psi2}: we plot $P(\omega)$ for two cases, one corresponding to subdiffusion, with a small value of $\epsilon$, and one to diffusion of momentum. In the first case (continuous line) a power-law behavior is observed, while in the second (dashed line) the power spectrum is flat in $\omega$. The dependence of the power law exponent of $P(\omega)$ on $K$ and $\epsilon$ is shown in FIG.~\ref{fig:plk411}, where we consider  $K=4.0$, panel (a), and $K=11.0$, panel (b). The dynamics is described by the full green squares which represent the power law exponent of the kinetic energy growth. The exponent of $P(\omega)$ vanishes (empty blue squares) when the dynamics is diffusive and it is negative when it is subdiffusive. 

The high frequency behavior of $P(\omega)$ is characterized by a series of peaks whose positions depend on $K$; by increasing  $\epsilon$ they spread and become smoother until they disappear when the system enters the diffusive region of the $(K,\epsilon)$ plane. In panel (b) of FIG.~\ref{fig:pow_spectr_psi2} this property is shown plotting the power spectrum for increasing values of $\epsilon$ at fixed $K$: we chose $\epsilon\leq 1$  it clearly appears that the peaks coincide in the three cases.

\subsubsection{Power spectrum $P_{ac}(\omega)$} 

\begin{figure*}
  \begin{center}
    \begin{tabular}{cc}
    \hspace{-0.5cm}\resizebox{9.0cm}{!}{\includegraphics{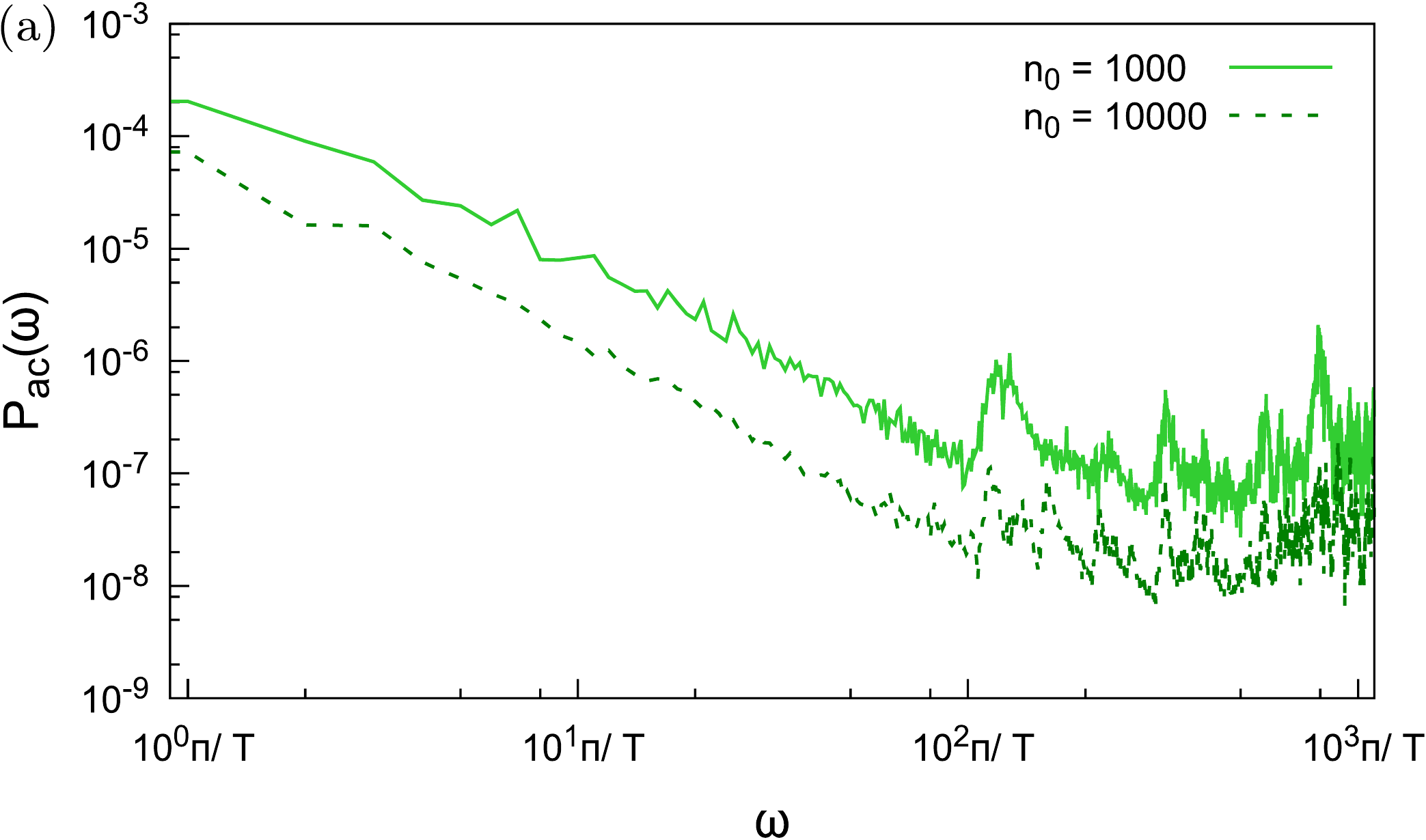}}&
    \hspace{0.5cm}\resizebox{9.0cm}{!}{\includegraphics{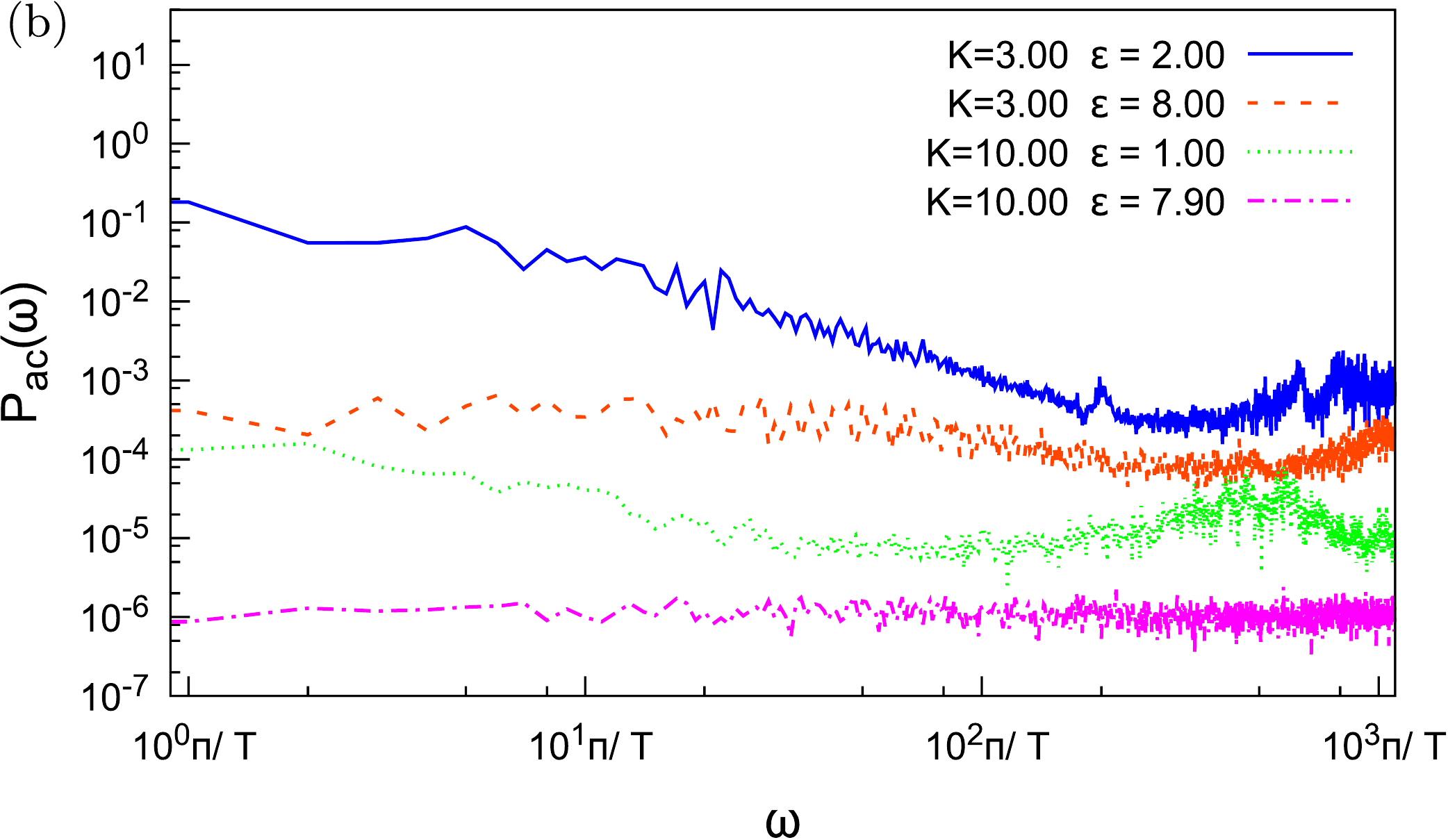}}
    \end{tabular}
  \end{center}
  \caption{(a) $P_{ac}(\omega; n_0)$ is plotted for two values of $n_0$: in the low frequencies region the slope of the curves is the same while the initial amplitude changes. This scaling is related to the power-law time dependence of $\sigma_{MF}$. Numerical parameters:  $K=6.0$ and $\epsilon = 0.52$. (b) A power-law behavior can be observed in the curve corresponding to $K=3.00$, $\epsilon=2.00$ (first curve from above) and it flattens as $K$ and $\epsilon$ are increased. The bottom line has been shifted down by an order of magnitude for a better visibility. In the simulations $\kbar= 2.89$.}
  \label{acf_diff_t_par}
\end{figure*}

Let us now discuss the power spectrum of the time-correlator $P_{ac}(\omega; n_0)$ at different $n_0$:  if the process $\psi_{MF}(n)$ is stationary,  the time-correlator $c(k;n_0)$ and its power spectrum are independent on $n_0$.
The small frequency results for our case are shown in panel (a) of FIG.~\ref{acf_diff_t_par} where we plot $P_{ac}(\omega; n_0)$ corresponding to $n_0=10^3$ and $n_0=10^4$: 
the two curves show a power-law behavior at low $\omega$ with the same exponent. The larger is $n_0$, however, the smaller the amplitude of $P_{ac}(\omega)$. It follows that $c_{MF}(k;n_0)$ scales to 0 as $n_0$ is increased: 
this result leads us to conclude that $\psi_{MF}(n)$ is not a stationary signal.
%
%

As already mentioned, $P_{ac}(\omega; n_0)$ decays like a power law in the subdiffusive region of the $(K, \epsilon)$ plane. 
This behavior is smoothed by increasing $\epsilon$ or $K$ until it disappears when diffusion starts: 
the exponent of the power law reduces and a uniform region at low $\omega$ appears. In the diffusive region of the $(K, \epsilon)$ plane $P_{ac}(\omega)$ is flat. 

In panel (b) of FIG.~\ref{acf_diff_t_par} we qualitatively show how $P_{ac}(\omega; n_0)$ changes as $\epsilon$ is increased. We consider $K=3.0$, for which the system is always in the subdiffusive region  (see FIG.~\ref{fig:quant_subdiff}), and $K=10.0$, for which the system passes from subdiffusive to diffusive as $\epsilon$ is increased. 
In the first case $P_{ac}(\omega)$ exhibits a power-law behavior at low frequencies for $\epsilon = 2.0$: this behavior is smoothed out when $\epsilon = 8.0$. 
In the second case the power-law behavior is much less evident when $\epsilon=1.0$; it completely disappears when $\epsilon=7.9$ and the system is diffusive. 
In FIG.~\ref{fig:plk411} the power law exponents relative to $P_{ac}(\omega)$ are plotted (red circles): for $K=4.0$, see panel (a), the exponent approaches the value of $-0.5$ without vanishing. On the other side in panel (b) we set $K=11.0$ and it vanishes for $\epsilon>4.0$: indeed for higher values of $\epsilon$ the system is diffusive.

At high frequencies $P_{ac}(\omega; n_0)$  is characterized by some peaks whose positions depend on $K$, similarly to what has been found for $P(\omega)$.
%
%

\subsubsection{Variance of the mean field parameter $\sigma_{MF}(n)$}

\begin{figure}
\hspace{0cm}\resizebox{\columnwidth}{!}{\includegraphics{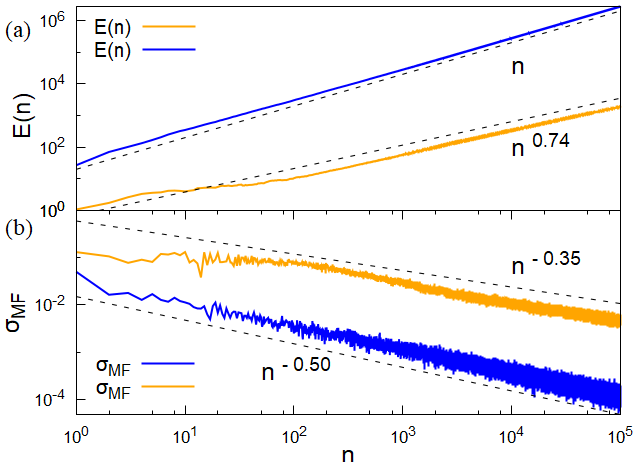}}
\caption{\label{fig:comp_diff_sub}Two cases of dynamics are considered, a diffusive one with $K=30.0, \ \epsilon = 4.90$ (blue, dark line, color online) and a subdiffusive one with $K=6.0, \ \epsilon = 0.52$ (orange, light one). (a) The growth of the kinetic energy is plotted for the two cases;  the dotted lines stresses the relative slopes. (b) The evolution of $\sigma_{MF}$ is plotted for the two cases so the corresponding power law behaviors are enhanced.}
\end{figure}

Let us now turn to $\sigma_{MF}(n)$ which is found to show a power law behavior, much clearer and robust than the one exhibited by $P_{ac}(\omega; n_0)$ and $P(\omega)$. In the subdiffusive region $\sigma_{MF}(n)$ decreases  as $ n^{-\beta}$, with $\beta$ slightly varying between to $0.3\div 0.4 $ while in the diffusive one $\sigma_{MF}(n)\sim n^{-\beta}$ with $\beta\simeq0.5$. In FIG.~\ref{fig:comp_diff_sub} $\sigma_{MF}(n)$ is plotted in two particular cases, one in the subdiffusive region and the other in the diffusive one. 

In panel (a) of FIG.~\ref{fig:plk411} we set $K=4.0$: the system is always subdiffsive and the power law exponent of  $\sigma_{MF}$ (orange triangles) is constant as $\epsilon$ is varied. In panel (b) we set $K=11.0$: the system passes from subdiffusive to diffusive for $\epsilon = 4.0$ as it can be seen in FIG. \ref{fig:quant_subdiff}. Accordingly, the exponent of  $\sigma_{MF}$ tends to $-0.5$; this transition is also enhanced by the exponent of the time-correlator (red circles) which vanishes at $\epsilon = 4.0$. 
 
Of course, the behavior of $\sigma_{MF}(n)$ clearly shows that $c_{MF}(k=0; n_0)$ is not stationary. 
In order to better understand how the features we are analyzing are relevant for the dynamics of our system we generate two artificial signals, $\phi(n)$ and $f(n)$, with some of the spectral properties we have found in $\psi_{MF}(n)$ and study the dynamics of a system perturbed by them instead of $\psi_{MF}(n)$.
The first signal $\phi(n)$ has a power spectrum like the one in FIG.~\ref{fig:pow_spectr_psi2} (power law behavior in panel (a)) and random phases assigned to the Fourier coefficients: the corresponding evolution operator, according to the definition in Eq.(\ref{quant_Uf2}), contains the kicking modulation $K'(n)=K(1-\epsilon \phi(n))$. 
The dynamics of this system is found to be subdiffusive up to a finite time, after which $E(n)$ grows linearly in time, analogously to what was found in the classical system in Ref.~\cite{PhysRevA.40.6130}: 
this means that the features of $P(\omega)$ are not a sufficient ingredient to reproduce the power law growth of $E(n)$.
On the other side, if
we  take $K''(n) = K(1-\epsilon f(n))$, where $f(n)$ is a stationary white noise process, the power law is uniform in $\omega$ and the energy grows linearly in time.
Therefore, while some features of the dynamics obtained can be associated to the properties of the time series the robust subdiffusion observed cannot be reproduced 
by a simple Gaussian process.


\section{Conclusions and perspectives}
In conclusion we have studied the ergodicity and energy absorption of a quantum chain of coupled kicked rotors. We have found a mapping of the $N$-body kicked rotor to a $N$-dimensional Anderson model in momentum space. This mapping has given us the possibility to make predictions on the energy dynamics of the kicked rotors: when $N>2$ there is a dynamical localization/delocalization transition which we have numerically observed in the energy dynamics and in the localization properties of the Floquet states in the momentum basis. 

Going to the thermodynamic limit $N\to\infty$ we find that the system is always dynamically delocalized. We have studied delocalization in this limit both in the corresponding Anderson model and directly in the coupled rotors model. In the first case, we have shown that the delocalization threshold vanishes; in the second we have used a mean field approach and found that the energy increases in a subdiffusive way in time. This is a genuine quantum phenomenon, since in the corresponding classical case the energy increases diffusively in time. This subdiffusion occurs together with some peculiar power-law behaviours of the mean-field order parameter, its Fourier transform and its time-correlator.
The effective mean field model suggests a comparison with other related models where there is a breaking of localization which can lead to subdiffusive processes. 
Examples of that are kicked rotors with a non-linear Hamiltonian or a modulated kicking and disordered lattice models with a nonlinearity in the Hamiltonian. 

 Our findings provide a clear example of many body driven dynamics where quantum mechanics qualitatively changes the regularity/ergodicity properties of the system with important consequences on energy absorption. This can be an important issue in the designing and working of quantum computers, as it already emerges from studies about quantum simulation of a single KR~\cite{PhysRevE.67.046220}. One perspective of future work is the application of our mapping on an Anderson model to other periodically driven models. A more ambitious one is the research of a driven system which can be mapped on a many body localized lattice model in momentum space. 

From the experimental point of view, the long-time coherent dynamics of Hamiltonians similar to ours can be realized in the framework of ultracold atoms in optical lattices~\cite{Bloch_RMP08,Bloch_Nat,Eckardt_RMP_17} and superconducting quantum circuits~\cite{Houck_Nat}. Although a pulsed field can be realized in single-particle models~\cite{PhysRevLett.75.4598}, pulsed interactions are not easy to engineer. Nevertheless, in the single rotor case the localization physics does not change when a sinusoidal driving is applied~\cite{Stockmann} and we expect the same result in the many coupled rotor case. Driven short range interactions can be engineered by means of Feschbach resonances in the ultracold atoms framework, and through SQUIDS in a time-dependent magnetic field in the case of superconducting circuits. Concerning driven long-range interactions, in principle they could be engineered using superconducting circuits of appropriate topology.
\acknowledgments
We acknowledge useful discussions with S.~Flach, M.~Fava, S.~Pappalardi and A.~Polkovnikov.  We acknowledge M.~Fava and E. G.~Dalla Torre for useful comments on the manuscript. R.~F. kindly acknowledges support from EU through project QUIC under grant agreement 641122, the National
Research Foundation of Singapore (CRP - QSYNC) and the Oxford Martin School. A.~R. acknowledges financial support from EU through project QUIC and from ``Progetti interni - Scuola Normale Superiore''. A.~R. dedicates this paper to the dear memory of his friend and mentor Ettore Montanari.
\appendix

\section{Classical interacting model} \label{classical:sec}
%
%

In this Appendix we discuss the behavior of the classical counterparts of the models defined in Eq. (\ref{H_sr:eqn}). We indicate the angle and momentum variables relative to the $i$ rotor at time $n$ as $\{\theta_i^n, p_i^n\}$.

A useful frame for understanding the dynamics of our models is provided by the seminal work of Nehkhoroshev (see Ref.~\cite{Nekhoroshev1971}) and other works (see Refs.~\cite{konishi, PhysRevA.40.6130}) about the classical dynamics of our system. For reader's convenience we review some known results and apply them to our models.

The kicked rotor model Hamiltonian can be written as 
\begin{equation}\label{eq:nek_1}
H=H_0(\theta, p)+K\, H_I(\theta, p; t)
\end{equation}
 where $H_0$ is an integrable Hamiltonian and $H_I$ breaks the integrability of the system with a strength given by $K$. It is relevant that both $H_0$ and $H_I$ are periodic in $\theta$.
For a system with two degrees of freedom, like the single rotor, we have seen in Section \ref{sec_sr} that for $K<K_c$ there are regions in the phase space in which the trajectories keep being closed (this result is in agreement with the KAM theory, as discussed in Section \ref{sec_sr}). The phase space is therefore divided in several regions by these trajectories and the dynamics of the system is not ergodic. The system exhibits, as already discussed, classical dynamical localization.

Nekhoroshev's theorem deals with the dynamics of a system with an Hamiltonian like the one defined in Eq. \eqref{eq:nek_1} but with more than two degrees of freedom. It states that, given an initial condition for the momentum variables $\{p_i^0\}_{1\leq i\leq N}$, one finds~\cite{konishi}
\begin{equation}\label{eqn:nek_2}
||{\bf p}^n-{\bf p}^0||<K^\alpha
\end{equation}
for $n<n^*$. We have $n^*\sim1/K\,\exp\{1/K^\beta\}$, $\beta\sim1/(\mathrm{polynomial\ function\ of}\ N)$ and $\alpha >0$. This is the same mechanism which allows the orbits of planets to remain stable in very long times: this should emphasize that if $K \ll 1$ the time during which condition in Eq. (\ref{eqn:nek_2}) is satisfied can be very long. After this time the trajectories of the system become unstable: their localization in the phase space is broken and the dynamics becomes ergodic~\cite{NOte14}.

For time independent Hamiltonians this means that the trajectories span the whole energy shell: averages can be computed using the micro-canonical ensemble.

For a time dependent Hamiltonian, the energy is not conserved and thus the trajectories will spread in all the phase space. 
This means that the system heats without a bound and thermalizes at $T=\infty$: 
this is indeed the case of our system, in which the kick breaks the integrability of the Hamiltonian. 

Now we numerically check this delocalization process for the two cases we are studying, the long range and the short range interacting ones.
Since we are interested in the dynamics at long times we choose amplitudes of the kick (namely the parameters $K$ and $\epsilon$ in Hamiltonian of Eq.\eqref{H_a:eqn}) for which the time $n^*$ is negligible.

We focus on the classical dynamics of Eq.~\eqref{H_a:eqn} in the case of infinite-range interactions.
It is possible to  integrate exactly the Hamilton equations for each rotor over a period, and obtain a map for the stroboscopic evolution of the system: restricting to discrete times $t_n=n$ we have
\begin{align}\label{class_map_p}
p_i^{n+1} &= p_i^n + K \bigg[ \sin \theta_i^{n} -\frac{\epsilon}{(N-1)} \sum_{j\neq i}\sin (\theta_i^{n}-\theta_{j}^{n})\bigg],   \\ \label{class_map_th}
\theta_i^{n+1} &= \theta_i^{n} + p_i^{n+1}.
\end{align}
We consider many realizations of the dynamics of the system sampling different initial conditions; they are chosen giving a uniformly random angle to each rotor and setting $p_i^0=0 \ \forall i=1...N$.
We focus on the time-evolution of the kinetic energy per rotor averaged over the ensemble of the initial conditions
\begin{equation}\label{class_kin_en}
  E(n) = \frac{1}{2N}\sum_{i=1}^N \overbar{(p_i^n)^2}\,.
\end{equation}
This quantity is proportional to the variance of the momenta distribution in time  and then gives information on the spreading in time of this distribution. 

In the numerical simulations that follow we set $N=100$: this number of rotors is sufficient to avoid boundary effects and simulate the $N\rightarrow \infty$ limit.
In the ergodic regime the time and space correlations between the angles of the rotors rapidly decay to zero (as it always occurs in chaos~\cite{Ott:book,Ruelle:book}): this implies in particular that 
\begin{align}\label{corr}
&\langle\cos (\theta_i^n-\theta_{j}^n)\cos (\theta_{i'}^m-\theta_{j'}^m)\rangle\\ \nonumber
&=\frac{1}{2}\delta_{n\,m}(\delta_{i\,i'}\delta_{j'\,j}+\delta_{i\,j'}\delta_{i'\,j}),
\end{align}
where the average is taken over the ensemble of the initial conditions.
Now we consider the following expression for the momentum at time $n$
\begin{equation}\label{class_p_ev}
p_i^{n} = K\sum_{\tau=0}^{n-1}\bigg[ \sin \theta_i^{\tau} -\frac{\epsilon}{(N-1)} \sum_{j\neq i}\sin (\theta_i^{\tau}-\theta_{j}^{\tau})\bigg].
\end{equation}
By squaring it and using Eq. \eqref{corr} we obtain the following coefficient describing the linear increase of the kinetic energy for the long-range interacting model:
\begin{equation}
D_{lr} = \frac{1}{4}K^2\left(1+\frac{\epsilon^2}{N-1}\right)\,.
\end{equation}

\begin{figure}
\hspace{-0.5cm}\resizebox{\columnwidth}{!}{\includegraphics{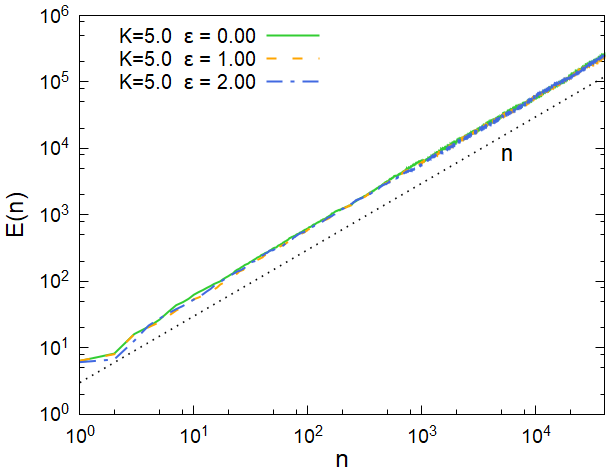}}
\caption{\label{fig:class_int_diff}The time evolution of $E(n)$ is plotted together with the curve $n$ as a guide to the eye to show the diffusive growth of the kinetic energy of the system. The diffusion coefficient is $D_{class}\sim K^2/4 = 6.25$: this is the expected value for the single kicked rotor and it is the same for all the values of $\epsilon$, since the curves are superposed. $N_{rotors}=100$ in the simulations.}
\end{figure}

Note that for $N\gg 1$ the diffusion coefficient coincides with the single-rotor one for $K>K_c$. 
In FIG.~\ref{fig:class_int_diff} this property is clearly shown: $E(n)$, computed at fixed $K=5.0$ but different values of $\epsilon=0, 0.5, 1.0$, always shows the same behaviour, growing linearly in time with the same angular coefficient. 

\begin{figure}
\hspace{-0.5cm}\resizebox{\columnwidth}{!}{\includegraphics{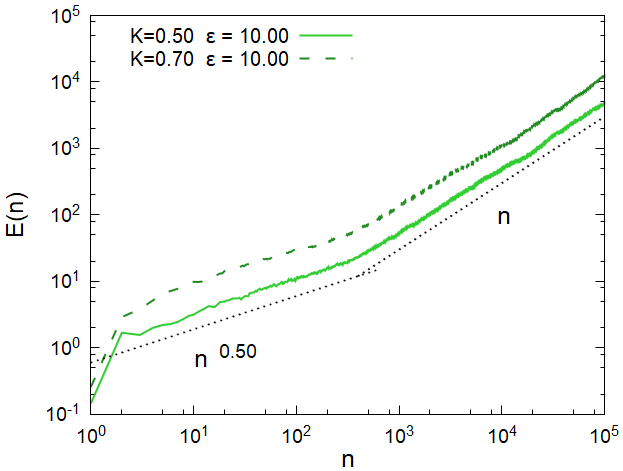}}
\caption{\label{fig:class_an_diff}The time evolution of $E(n)$ is plotted;  the two  regimes, the subdiffusive and diffusive one, are clearly distinguishable in the log-log scale. $N_{rotors}=100$ in the simulations.}
\end{figure}

A remarkable difference emerges for $K<K_c$: in this case the single rotor manifests dynamical localization, while the presence of an interaction induces a growth of the kinetic energy which starts being only subdiffusive and becomes diffusive after a transient (see Ref.~\cite{PhysRevA.40.6130} for the same phenomenon in a different model). This behaviour is perfectly consistent with the Nekhoroshev theorem~\cite{konishi} and we show it in FIG.~\ref{fig:class_an_diff} by plotting the evolution of $E(n)$ for different values of $K$.

\begin{figure}
\hspace{-0.5cm}\resizebox{\columnwidth}{!}{\includegraphics{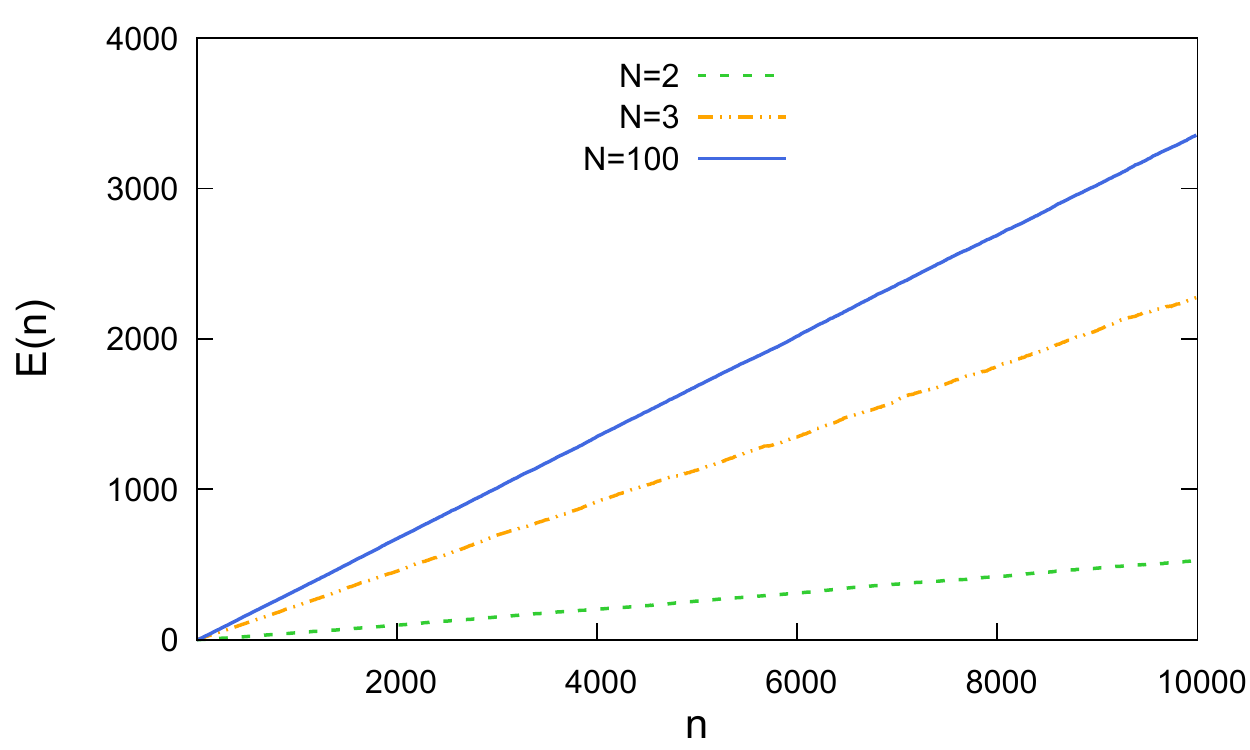}}
\caption{\label{fig:2_2_rot_class_diff}In this figure the time evolution of $E(n)$ is plotted with $K=0.5$ and $\epsilon=-2.0$. As the number of rotors is increased the diffusion coefficient grows approaching the asymptotic value ($N=100$ rotors). Nevertheless the exact value of the diffusion coefficient in Eq. \ref{eqn:nn_diff_coeff} is reached for $K\gtrsim2$.}
\end{figure}

The classical behavior of the short-ranged model is very similar. With analogous calculations we find that the diffusion coefficient for the kinetic energy, in absence of correlations and with $N\rightarrow\infty$, is
\begin{equation}\label{eqn:nn_diff_coeff}
D_{sr} = \frac{1}{4}K^2\left(1+\frac{\epsilon^2}{2}\right)\,.
\end{equation}
With small $N$ (like the cases $N=2,3$ we consider in the text for the quantum model) finite size effects~\cite{konishi} reduce the diffusion coefficient as it is shown in FIG. \ref{fig:2_2_rot_class_diff}. Moreover, some corrections due to correlations  modify the diffusion coefficient, as it has been seen in Ref.~\cite{2rot_1} for two rotors; they disappear for $K\epsilon\gtrsim 2$.

To conclude, we have shown that a classical interacting rotors model exhibits an ergodic behavior and, at long times, a linear growth of the kinetic energy: this characteristic is manifested for all the values of $K$ and $\epsilon$, independently from the number of rotors.

\section{Exactness of the mean field approximation for $N\to\infty$}\label{mf_calc}

In this Appendix we demonstrate that $\langle\hat\delta_r\hat\delta_s\rangle\rightarrow0$ in the limit $N\rightarrow\infty$ if we start from a separable state at time $n=0$.
We first observe that $\langle\hat\delta_r\hat\delta_s\rangle=c(r, s; n)$, where
\begin{align}\nonumber
&c(r, s; n)= \\ \label{quant_corr}
&\langle\cos\hat\theta_r(n)\cos\hat\theta_s(n)\rangle-\langle\cos\hat\theta_r(n)\rangle\langle\cos\hat\theta_s(n)\rangle.
\end{align}
is the time dependent, spatial connected correlator between different rotors. Therefore the relation $\langle\hat\delta_r\hat\delta_s\rangle\rightarrow0$ for $N\rightarrow\infty$ means that the system does not develop spatial correlation during the evolution in the thermodynamic limit. 

We set $\kbar=1$; for simplicity we define $\alpha=\epsilon K /(N-1)$ and then we set $K=0$: without losing generality  we are considering only the interacting part in the kick. 

The scheme of the demonstration is the following: we expand the term $\langle\cos\hat\theta_r(n)\cos\hat\theta_s(n)\rangle$ keeping the ones which are $O(1/N)$. Some of the resulting terms are canceled out by $\langle\cos\hat\theta_r(n)\rangle\langle\cos\hat\theta_s(n)\rangle$: we show that only a finite number of terms $O(1/N)$ remains: therefore, they vanish in the thermodynamic limit. 

Once we have demonstrated the absence of spatial correlations we define the operator $\hat\Delta_N=1/(N-1)\sum_{r\neq s}\hat\delta_r\hat\delta_s$: it is the sum of fluctuation terms which appears in Eq.~\eqref{eqn:formulona}. By applying the central limit theorem we show that it increases in a non extensive way. Therefore, it brings negligible contributions to the Hamiltonian in the thermodynamic limit.

As a first step we introduce some notation useful for the demonstration: we write the one-period propagator as $\hat U=\hat K\,\hat T$, with
$\hat K$ containing the kick part of the propagator and $\hat T$ the kinetic one.
We define $\hat K_r= \nep^{i\alpha\sum_{s\neq r}\cos(\hat\theta_r-\hat\theta_s)}$ and $\hat T_r=  \nep^{i\frac{\hat p_r^2}{2}}$.
We have $[\hat T_r, \hat T_s]=[\hat K_r, \hat K_s] = 0\ \forall\, r,s$; also,
\begin{equation}\label{eqn:commKTsingle}
\mathcal{C}_{1,2}(\hat \theta_r, \hat\theta_s, \hat p_r) =\pm[\nep^{\mp i\alpha\cos(\hat\theta_r-\hat\theta_s)}, \nep^{\mp i\frac{\hat p_r^2}{2}}]\,.
\end{equation}
These commutators are bounded operators, since they come from unitary operators. Moreover, one can easily deduce  that $||\mathcal{C}_{1,2}(\hat \theta_r, \hat\theta_s, \hat p_r)||\sim 1/(N-1)$ from a first order expansion of $\nep^{\mp i\alpha\cos(\hat\theta_r-\hat\theta_s)}$.
In the next subsection we go through the expansion of the term $\langle\cos\hat\theta_r(n)\cos\hat\theta_s(n)\rangle$.

\subsection{Expansion of $\langle\cos\hat\theta_r(n)\cos\hat\theta_s(n)\rangle$}
Given the initial factorized state $|\Psi_0\rangle$ 
the expectation value of the product at time $n=2$ is given by
\begin{align} \label{eqn:sqbra}
&\langle\Psi_0|\left[(U^\dagger)^2\cos\hat\theta_r\,U^2\right]\left[(U^\dagger)^2\cos\hat\theta_s\,U^2\right]|\Psi_0\rangle\,.
\end{align}
The choice of $n=2$ is motivated by the fact that at this time correlations start to develop.
We focus on the content of the left squared brackets. 
First, we write $\hat U$ and $\hat U^\dagger$ by using the definition given above. Then we simplify all the terms which freely commute and what remains is the following:
\begin{equation}
\hat T^{\dagger}\,\hat K_r^\dagger\, \hat T_r^\dagger\, \cos\hat\theta_r\, \hat T_r\,\hat K_r \,\hat T
\end{equation}
The kinetic operator $\hat T^{\dagger}$ on the right automatically simplifies with the relative term $\hat T$ in the right squared brackets so we neglect it. 
Also, we define $|\tilde\Psi_0\rangle=\hat T |\Psi_0\rangle$ so we can restrict our study to the term $\hat K_r^\dagger\, \hat T_r^\dagger\, \cos\hat\theta_r\, \hat T_r\,\hat K_r$. 
In order to expand this term we need to invert the operators $K_r^\dagger\, \hat T_r^\dagger$ and $ \hat T_r\,\hat K_r$ respectively, so we need to compute the two commutators $[\hat K_r^\dagger, \hat T_r^\dagger]$ and $[ \hat T_r,\hat K_r]$.

For the first commutator we have (we consider $r=N$ for simplicity but the generalization is straightforward):
\begin{align}\label{eqn:commKT}
&[\hat K_N^\dagger, \hat T_N^\dagger] =\mathcal{C}_1(\hat \theta_N, \hat\theta_n, \hat p_N)\prod_{\nu=1}^{N-2}\nep^{-i\alpha\cos(\hat\theta_N-\hat\theta_{\nu})}+\\ \nonumber
& \sum_{n=1}^{N-2}\bigg[\left(\prod_{\nu=1}^{n}\nep^{-i\alpha\cos(\hat\theta_N-\hat\theta_{\nu})}\right)\mathcal{C}_1(\hat \theta_N, \hat\theta_n, \hat p_N) \\ \nonumber
&\left(\prod_{\mu=n+1}^{N-2}\nep^{-i\alpha\cos(\hat\theta_N-\hat\theta_{\mu})}\right)\bigg]\,.
\end{align} 

Also, for each term in the sum labeled by $n$ we have:
\begin{align}
&\left(\prod_{\nu=1}^{n}\nep^{-i\alpha\cos(\hat\theta_N-\hat\theta_{\nu})}\right)\mathcal{C}_1(\hat \theta_N, \hat\theta_n, \hat p_N) = \\ \nonumber
&\left(\prod_{\nu=1}^{n-1}\nep^{-i\alpha\cos(\hat\theta_N-\hat\theta_{\nu})}\right)\bigg(\mathcal{C}_1(\hat \theta_N, \hat\theta_n, \hat p_N) \nep^{-i\alpha\cos(\hat\theta_N-\hat\theta_{n})}+\\ \nonumber
&\xi_1(\hat \theta_N, \hat\theta_n, \hat p_N) = \dots \\ \nonumber
&=\mathcal{C}_1(\hat \theta_N, \hat\theta_n, \hat p_N) \left(\prod_{\nu=1}^{n}\nep^{-i\alpha\cos(\hat\theta_N-\hat\theta_{\nu})}\right) + \\ \nonumber
& \sum_{\nu=1}^n\left(\xi_1(\hat \theta_N, \hat\theta_\nu, \hat p_N) \prod_{\mu\neq\nu}^{n}\nep^{-i\alpha\cos(\hat\theta_N-\hat\theta_{\mu})}\right)\\ \nonumber
&+ O(1/N^2),
\end{align}
where 
\begin{equation}\label{eqn:defxi}
\xi_{1,2}(\hat \theta_r, \hat\theta_s, \hat p_r) =\pm[\nep^{\mp i\alpha\cos(\hat\theta_r-\hat\theta_s)}, \mathcal{C}_{1,2}]\,.
\end{equation}
Note that $\xi_{1,2}$ is an operator whose norm is $O(1/N^2)$. The $O(1/N^2)$ terms in the last equation come from higher order commutators and we henceforth neglect them.
Therefore Eq.~\eqref{eqn:commKT} can be rewritten as follows:
\begin{align}\label{eqn:commKTbis}
&[\hat K_N^\dagger, \hat T_N^\dagger] \simeq\\ \nonumber
& \sum_{n=1}^{N-1}\left[\mathcal{C}_1(\hat \theta_N, \hat\theta_n, \hat p_N)+\sum_{\nu=1}^{n}\xi_1(\hat \theta_N, \hat\theta_\nu, \hat p_N)\right]\hat K^\dagger_N,
\end{align}
An analogous result can be obtained for the commutator $[\hat T_N, \hat K_N]$:
\begin{align}\label{eqn:commKTbis2}
&[\hat T_N, \hat K_N] \simeq\\ \nonumber
&\hat K_N\, \sum_{n=1}^{N-1}\left[\mathcal{C}_2(\hat \theta_N, \hat\theta_n, \hat p_N)+\sum_{\nu=1}^{n}\xi_2(\hat \theta_N, \hat\theta_\nu, \hat p_N)\right].
\end{align} 
%
The important point of Equations~\eqref{eqn:commKTbis} and~\eqref{eqn:commKTbis2} is that the commutators can be written as the sum of $N-1$ terms of order $1/(N)$ and $(N-1)^2/2$ terms of order $O(1/N^2)$, up to higher order terms.


Now we can factorize the term $\hat K_r^\dagger\, \hat T_r^\dagger\, \cos\hat\theta_r\, \hat T_r\,\hat K_r$ by  using Equations ~\eqref{eqn:commKTbis} and~\eqref{eqn:commKTbis2}:
\begin{align}\label{eqn:cosleft}
&\hat K_r^\dagger\, \hat T_r^\dagger\, \cos\hat\theta_r\, \hat T_r\,\hat K_r \\ \nonumber
=&\left[\left(\hat T_r^\dagger + \sum_{n\neq r}\mathcal{C}_1(\hat \theta_r, \hat\theta_n, \hat p_r)+
\sum_{n\neq r\,,\nu=1}^{n} \xi_1(\hat \theta_r, \hat\theta_\nu, \hat p_r)\right)
\hat K^\dagger_r \right]\\ \nonumber
& \cos\hat\theta_r \\ \nonumber
&\left[ \hat K_r \,\left(\hat T_r^\dagger + \sum_{m\neq r}\mathcal{C}_2(\hat \theta_r, \hat\theta_m, \hat p_r)+
\sum_{n\neq r\,,\nu=1}^{n} \xi_2(\hat \theta_r, \hat\theta_\nu, \hat p_r)\right) \right] \\ \nonumber
=&\, \hat T_r^\dagger\, \cos\hat\theta_r\,  \hat T_r + \sum_{n\neq r}\mathcal{C}_1(\hat \theta_r, \hat\theta_n, \hat p_r) \cos\hat\theta_r \\ \nonumber
&+ \cos\hat\theta_r  \sum_{m\neq r}\mathcal{C}_2(\hat \theta_r, \hat\theta_m, \hat p_r) +\sum_{n\neq r\,,\nu=1}^n \xi_1(\hat \theta_r, \hat\theta_\nu, \hat p_r) \cos\hat\theta_r \\ \nonumber
& +\cos\hat\theta_r  \sum_{n\neq r\,,\nu=1}^n \xi_2(\hat \theta_r, \hat\theta_\nu, \hat p_r).
\end{align}
Analogously the right squared brackets term in Eq.~\eqref{eqn:sqbra} returns:
\begin{align} \label{eqn:cosright}
&\hat K_s^\dagger\, \hat T_s^\dagger\, \cos\hat\theta_s\, \hat T_s\,\hat K_s \\ \nonumber
=&\, \hat T_s^\dagger\, \cos\hat\theta_s\,  \hat T_s + \sum_{n\neq s}\mathcal{C}_1(\hat \theta_s, \hat\theta_n, \hat p_s) \cos\hat\theta_s +\\ \nonumber
& \cos\hat\theta_s  \sum_{m\neq s}\mathcal{C}_2(\hat \theta_s, \hat\theta_m, \hat p_s) +\sum_{n\neq s\,,\nu=1}^n \xi_1(\hat \theta_s, \hat\theta_\nu, \hat p_s) \cos\hat\theta_s \\ \nonumber
& +\cos\hat\theta_s  \sum_{n\neq s\,,\nu=1}^n \xi_2(\hat \theta_s, \hat\theta_\nu, \hat p_s).
\end{align}
Now we multiply the results in Equations~\eqref{eqn:cosleft} and~\eqref{eqn:cosright} keeping explicit only first order terms and take the expectation values. At the zeroth order we have 
\begin{equation}
\langle\tilde\Psi_0| \hat T_r^\dagger\, \cos\hat\theta_r\,  \hat T_r \hat T_s^\dagger\, \cos\hat\theta_s\,  \hat T_s |\tilde\Psi_0\rangle\,,
\end{equation}
which represents the evolution without kick and can be factorized. 
At the first order we have two sums which provide respectively
\begin{align}
&\Sigma_r = \langle\tilde\Psi_0| \hat T_r^\dagger\, \cos\hat\theta_r\,  \hat T_r \\ \nonumber
& \sum_{n\neq s}\left[\mathcal{C}_1(\hat \theta_s, \hat\theta_n, \hat p_s )\cos\hat\theta_s +\cos\hat\theta_s\mathcal{C}_2(\hat \theta_s, \hat\theta_n, \hat p_s)  \right]|\tilde\Psi_0\rangle\ 
\end{align}
and an analogous term $\Sigma_s$ is defined. 
In an analogous way the sums of the second order terms $\Pi_{1,2}$ containing $\xi_{1,2}$ must be considered.
\subsection{$O(1/N)$ terms in $c(r, s; n)$}

We henceforth explain how the extensive sums $\Sigma_{r,s}$ and $\Pi_{1,2}$ reduce to a non extensive amount of contributions in $c(r, s; n)$. We explain the mechanism for $\Sigma_{r,s}$ but it equally apply for the $\Pi_{1,2}$.

We indeed focus on $c(r, s; n)$ and check which terms does not cancel out when we take the difference $\langle\cos\hat\theta_r(n) \,\cos\hat\theta_s(n)\rangle-\langle\cos\hat\theta_r(n)\rangle\langle\cos\hat\theta_s(n)\rangle$. 
Almost all the terms in $\Sigma_{r,s}$ can be factorized and therefore cancel out with equal contributions coming from the product $\langle\cos\hat\theta_r(n)\rangle\langle\cos\hat\theta_s(n)\rangle$: the only exceptions are two terms with $n=r$ in $\Sigma_r$ and two with $n=s$ in $\Sigma_s$. 
Indeed we obtain four differences which do not cancel, one of those is
\begin{align}
&\langle\tilde\Psi_0| \hat T_r^\dagger\, \cos\hat\theta_r\,  \hat T_r\ \mathcal{C}_1(\hat \theta_s, \hat\theta_r, \hat p_s )\cos\hat\theta_s|\tilde\Psi_0\rangle\ - \\ \nonumber
&\langle\tilde\Psi_0| \hat T_r^\dagger\, \cos\hat\theta_r\,  \hat T_r|\tilde\Psi_0\rangle\ \langle\tilde\Psi_0| \mathcal{C}_1(\hat \theta_s, \hat\theta_r, \hat p_s )\cos\hat\theta_s|\tilde\Psi_0\rangle\,.
\end{align}
The three others  terms have the same structure. 

Therefore $c(r, s; n=2)$ does not vanish because of a finite number of $O(1/N)$ corrections (now we neglect the $O(1/N^2)$ terms coming from $\xi_{1,2}$): we have found  the first  contributions in Equations~\eqref{eqn:cosleft} and~\eqref{eqn:cosright}. Also, we have found the contributions $O(1/N)$ coming from $\Sigma_r$ and $\Sigma_s$. At a time $n>2$ the number of these contributions linearly increases, although it is always finite. Since, anyway, the limit $N\rightarrow\infty$ is taken before the evolution starts, the correlations are always going to zero like $1/N$: the final result we obtain is that  $\langle\hat\delta_r\hat\delta_s\rangle\rightarrow\langle\hat\delta_r\rangle\langle\hat\delta_s\rangle$ in the thermodynamic limit.

\subsection{Central limit theorem and conclusion}
According to the previous result, we concentrate on the operator $\hat\Delta_N$ defined above. Since $\langle\hat\delta_r\hat\delta_s\rangle=\langle\hat\delta_r\rangle\langle\hat\delta_s\rangle$ its expectation value over the state  $|\phi(n)\rangle$ is:
\begin{align} \label{strong_conv}
&\langle\phi(n)|\hat\Delta_N|\phi(n)\rangle = \\ \nonumber
& \frac{1}{(N-1)} \sum_{i\neq j}\bigg[(\langle\cos\hat\theta_i \rangle-\chi)  (\langle\cos\hat\theta_j \rangle-\chi) )\bigg]
\end{align}
Each of the two sums represents the fluctuations of a set of independent, random variables: we can apply the central limit theorem and state that $1/(N-1)\sum_{i\neq j}(\langle\cos\hat\theta_i \rangle-\chi)\,\sim 1/\sqrt{N}$. 
It follows that $\langle\phi(n)|\hat\Delta_N|\phi(n)\rangle \sim \sqrt{N}$: since the fluctuation term in the Hamiltonian grows less than extensively it can be neglected in the thermodynamic limit. We conclude that the mean field approach is therefore exact.


\begin{thebibliography}{100}
\bibitem{Note_sig}
\newblock{The dynamics preserves the symmetry $p\to-p$, therefore if we choose all the initial conditions of the ensemble such that $p(0)=0$ we have $\overline{p}(n)=0$ $\forall\,n$ and then $\sigma_p^2(n)\equiv\overline{{p^2}(n)}$.}
\bibitem{NOte10}
This phenomenon is observed for every value of $K$, with $\kbar\neq 2 m \pi$
  with $m\in \mathbb{Z}$: for these values, instead, quantum resonances occur
  and kinetic energy $E(n)$ grows quadratically in time.
\bibitem{NOte2}
Eq.~\eqref{evoluzione:eqn} can be easily derived considering that
  Eq.~\eqref{eigenstates:eqn} implies
  $\hat{U}=\sum_{\alpha=-\infty}^\infty\nep^{-i\mu_\alpha}\ket{\phi_\alpha}\bra{\phi_\alpha}$.
\bibitem{NOte4}
Assuming that the wave-function exponentially decays $R_\alpha({\bf
  m})\simeq\mathcal{N}\nep^{-|\mathbf{m}-\mathbf{m}_\alpha|/\lambda}$ and
  approximating the sums over ${\bf m}$ as integrals, we find that the
  normalization factor is
  $\mathcal{N}=\frac{1}{\sqrt{S_{N-1}(N-1)!\,(\lambda/2)^N}}$. Using this
  formula in Eq.~\eqref{timeav:eqn} and still approximating the sums as
  integrals, we get Eq.~\eqref{aver2:eqn}.
\bibitem{NOte3}
Pseudo randomness of the potential is achieved taking $\kbar / \pi$ incommensurate.
The sequence $\cos(\pi\alpha n^2)$, with $\alpha$ irrational,  has been already considered as a lattice pseudo random potential \cite{BrennerFishman} and it has been shown that, although it fails some empiric randomness tests, the localization properties of the related system are not different from the truly random Anderson one.
\bibitem{Note_Davide}
For $N=3$ we need to use TEBD algorithm on a truncated MPS representation of the system wavefunction, because the full Hilbert space is too large to be handled with exact methods (taking $M=17$, one already reaches a huge dimension of $(2M+1)^N=42875$). In the TEBD implementation performed here, we always force the truncation error to be less than $10^{-10}$, so to keep the bond-link dimension to a manageable size.
\bibitem{NOte7}
In the classical counterpart of our model we see an ergodic behaviour for any
  value of the kicking parameters -- see Ref.~\cite{konishi} and
  Appendix~\ref{classical:sec}. Even if not all the phase space is ergodic, the
  regular portion of it is too small to block diffusion of the
  trajectories~\cite{Arnold_Avez:book}.
\bibitem{NOte6}
The evolution operator defined in Eq. \eqref{quant_Uf2} preserves the even parity of the state  during its evolution: As a conclusion, if $\psi_{MF}(0)$ is real $\psi_{MF}(n)$ is real $\forall n$.
\bibitem{NOte14}
Trajectories which do not obey ergodicity can persist also at times $n>n^*$ But
  they are confined in a $N$ dimensional manifolds of the $2N-1$ dimensional
  energy shell: therefore when $N>1$ they do not create separated regions in
  the phase space and the dynamics of all the other trajectories is ergodic.
\end{thebibliography}

\providecommand{\noopsort}[1]{}\providecommand{\singleletter}[1]{#1}%

\end{document}